\documentclass[11pt,twocolumn]{article}
\usepackage[a4paper,left=1.8cm,right=1.8cm,top=2cm,bottom=2.2cm]{geometry}
\usepackage{graphicx}
\usepackage{amsmath,amssymb}
\usepackage{microtype}
\usepackage[utf8]{inputenc}

\usepackage{booktabs}
\usepackage{multirow}
\newcommand{\vek}[1]{\mathchoice{\displaystyle\boldsymbol#1}
{\textstyle\boldsymbol#1}{\scriptstyle\boldsymbol#1}
{\scriptscriptstyle\boldsymbol#1}}
\newcommand{\mat}[1]{\mathchoice{\displaystyle\mathbf#1}
{\textstyle\mathbf#1}{\scriptstyle\mathbf#1}
{\scriptscriptstyle\mathbf#1}}
\usepackage{dblfloatfix}
\usepackage{xcolor}
\newcommand{\R}[1]{#1}   

\begin{document}
\twocolumn[
\begin{center}

{\LARGE Identification of Random Material Properties\\[0.2em]
as Stochastic Inversion Problem}

\vspace{1em}

Eli\v{s}ka Ko\v{c}kov\'a$^{*}$, Anna Ku\v{c}erov\'a

\vspace{0.5em}

{\small
Faculty of Civil Engineering, Czech Technical University in Prague,\\
Th\'akurova 7, 166 29 Prague, Czech Republic
}

\vspace{0.5em}

{\small $^{*}$Corresponding author: eliska.kockova@fsv.cvut.cz}

\vspace{1em}

\textit{This is a preprint of a paper accepted for publication in Acta Polytechnica.}

\end{center}

\vspace{1em}

\noindent\textbf{Abstract:}
Heterogeneity of many building materials complicates numerical modelling of structural behaviour. The material randomicity can be manifested by different values of material parameters of each material specimen. To capture inherent variability of heterogeneous materials, the model parameters describing the material properties are considered as random variables and their identification consists in solving a~stochastic inversion problem. The stochastic inversion is based on searching for probabilistic description of model parameters which provides the distribution of the model response corresponding to the distribution of the observed data. The paper presents two different formulations of the stochastic inversion problem. The first formulation arises from the Bayesian inference of uncertain statistical moments of a prescribed parameters' distribution while the main idea of the second one utilizes nonlinear transformation of random model parameters from distribution of the observed data. 

\vspace{1em}


\vspace{1em}

]

\bigskip
\section{Introduction}
Randomness of material properties at different scales caused by their natural heterogeneity or imperfections created during manufacturing processes can significantly affect the structural behaviour. Developing and calibrating numerical models of systems involving heterogeneous materials requires not only accurate model predictions but also a comprehensive understanding of the associated uncertainties. Heterogeneity of a~material can be described in different ways depending on the real system to be modelled. In order to be more specific, we distinguish two principal categories of systems which fundamentally differ in involving randomness of heterogeneous material characteristics and thereby in identifying the underlying model parameters:
\begin{itemize}
    \item [] {\bf Deterministic system:} One object -- a specific laboratory specimen or more often a specific real-world structure - is called here a deterministic system. The object has some given yet unknown material properties in every point of its spatial domain. The relation between deterministic model inputs $\vek{x}$ and outputs $\vek{y}$ is given by a numerical model
    \begin{equation}
        \vek{y} = g(\vek{x}).
    \label{eq:model_deterministic}
    \end{equation}
    \item [] {\bf Stochastic system:} An ensemble of objects --- material samples or system realizations sharing certain common features that define the system, while other properties vary randomly from one instance to another.
    For example, a set of laboratory specimens, such as heterogeneous concrete cylinders of the same size made of the same mix of the concrete, where the particular samples differ in their morphology. The numerical model defines the relation between random model inputs $\vek{X}$ and random outputs $\vek{Y}$ as
    \begin{equation}
        \vek{Y} = g(\vek{X}).
    \label{eq:model}
    \end{equation}
\end{itemize}

The presented classification does not stem solely from the inherent properties of a given system; the distinction between deterministic and stochastic systems represents rather a pragmatic choice made by modellers. 

In this work, we focus on stochastic modelling of heterogeneous materials, where the inherent variability is represented by introducing randomness into the model input parameters in order to account for all relevant uncertainties, particularly the irregular spatial arrangement of the microstructure. Calibrating such a material model requires identification of its parameters' probability distribution function (PDF), which is referred to as a stochastic inversion problem.  Stochastic inversion enables to infer the unknown PDF of the model parameters from indirect experimental measurements. 

The experimental measurement is generally repeated several times. For a particular experimental realization $i$, these parameters have specific, fixed yet unknown, values $\vek{x}_i$. For an ensemble of $n$ specimens from the same heterogeneous material, the values $\vek{x}_i$ for $i=1,...,n$ are subject to unknown variability which aims to be identified.

The basic idea of probabilistic formulation of parameter identification is assuming the observable data $\vek{Z}$, the model response $\vek{Y}$ and the model parameters $\vek{X}$ as continuous random variables with their PDFs
\begin{equation}
    \begin{tabular}{l}
       $ \vek{Z} \sim f_{\vek{Z}}(\cdot)$ ,\\
       $ \vek{Y} \sim f_{\vek{Y}}(\cdot)$ ,\\
       $ \vek{X} \sim f_{\vek{X}}(\cdot)$ .\\
    \end{tabular}
\label{eq:RV}
\end{equation}
The solution of stochastic inversion for the numerical model (\ref{eq:model}) is the model input parameters' PDF $f_{\vek{X}}(\cdot)$ which provides the distribution of the model response $f_{\vek{Y}}(g(\vek{X}))$ corresponding to the distribution of the observed data $f_{\vek{Z}}(\cdot)$.

The material properties of a material sample (e.g. one concrete cylinder in the deterministic case) are typically characterised by a set material parameters \R{$\vek{x}=(x_1,...,x_{n_x})$} such as Young's modulus, Poisson's ratio etc. Material properties of a material (e.g. concrete B30 in the stochastic formulation) are represented by a random vector $\vek{X}=(X_1,...,X_{n_X})$. Experi\-ments often result in time-dependent data discretised into $n_t$ time steps. The observation and correspon\-ding numerical model are following in the deterministic case
\begin{equation}
    \begin{tabular}{cc}
       $ \vek{y} = (y_1,...,y_{n_t})$, & $y_t = g_t(\vek{x}) \text{ for } t = 1,...,n_t$ ,\\
    \end{tabular}
\label{eq:model_deterministic_time}
\end{equation}
and in the stochastic case
\begin{equation}
    \begin{tabular}{cc}
       $ \vek{Y} = (Y_1,...,Y_{n_t})$, & $Y_t = g_t(\vek{X}) \text{ for } t = 1,...,n_t$ .\\
    \end{tabular}
\label{eq:model_time}
\end{equation}

In order to make the inverse analysis successful, a proper distinction of occurring uncertainties is required. General classification  distinguishes two essential types of uncertainties \cite{Oberkampf:2002, DerKiureghian:2009:SS}:
\begin{itemize}
    \item [] {\bf Epistemic} uncertainty expresses a lack of knowledge, caused by e.g. measurement errors or a small number of measurements. This uncertainty can be reduced by any new information about the quantity of interest. In this paper, independent normally distributed experimental errors $\vek{E}~\sim~N(\vek{0},\vek{\sigma}_E^2\mat{I}$) are considered, the observations from the deterministic system become random
    \begin{equation}
        Z_t = g_t(\vek{x})+E_t \text{ for } t = 1,...,n_t .
    \label{eq:epistemic}
    \end{equation}
    \item [] {\bf Aleatory} uncertainty is irreducible stochasticity or simply variability of quantity of interest. It represents an inherent randomness which originates from modelling an ensemble of measurements e.g. from different locations or times as a random variable. The observations are random due to stochastic setup of the system with random inputs $\vek{X}$
    \begin{equation}
        Z_t = g_t(\vek{X}) \text{ for } t = 1,...,n_t .
    \label{eq:aleatoric}
    \end{equation}
\end{itemize}

\begin{figure*}
\begin{center}
\begin{tabular}{cc}
Deterministic system & Stochastic system \\
\toprule
\includegraphics*[width=75mm,keepaspectratio]{ 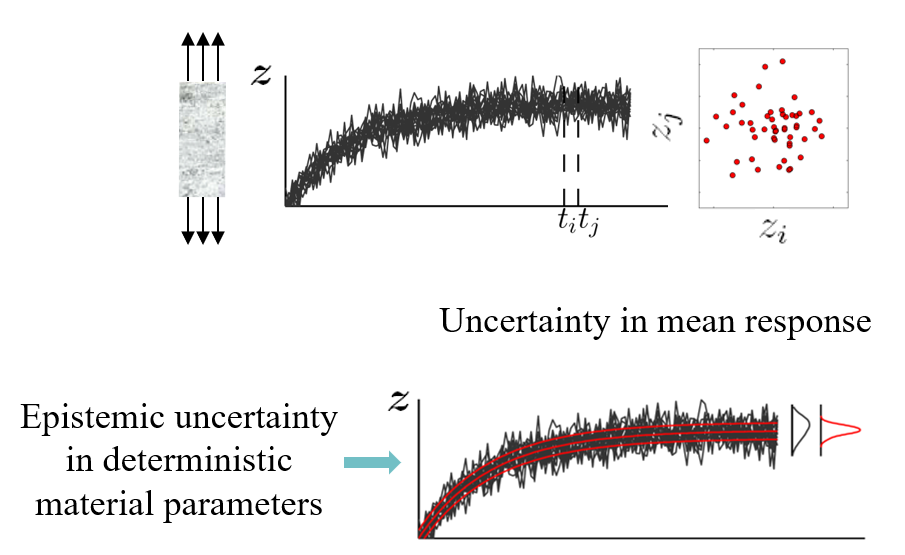} &
\includegraphics*[width=75mm,keepaspectratio]{ 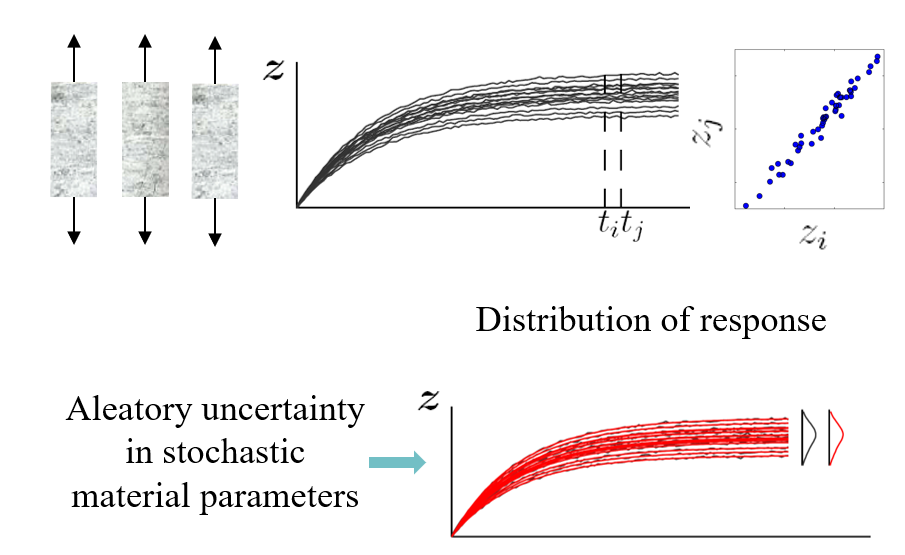} \\
$Z_t = g_t({\vek{x}}) + E_t$ & $Z_t = g_t({\vek{X}}) + E_t$\\
\midrule
     (a) &  (b)
\end{tabular}
\end{center}
\caption{Experimental outputs considered as random variables $Z_t$ due to (a) experimental error $E_t$ (one specimen with deterministic vector of material parameters ${\vek{x}}$) and (b) experimental error $E_t$ and stochastic material parameters characterized by random variables ${\vek{X}}$ (ensemble of specimens).} 
\label{fig:det-stoch}
\end{figure*}

We formulate a stochastic inversion problem invol\-ving both---epistemic and aleatoric--- uncertainties as follows
\begin{equation}
    \begin{tabular}{c}
        $ Z_t = g_t(\vek{X})+E_t \text{ for } t = 1,...,n_t$ ,\\
        $ \vek{X} \sim f_{\vek{X}}(\cdot)$ ,\\
        $ Z_t \sim f_{Z_t}(\cdot)  \text{ for } t = 1,...,n_t$ ,\\
        $ Y_t = g_t(\vek{X}) \sim f_{Y_t}(\cdot)  \text{ for } t = 1,...,n_t$ ,\\
        $ E_t \sim N(0,\sigma_E^2)  \text{ for } t = 1,...,n_t$ .\\
    \end{tabular}
\label{eq:stochastic_problem}
\end{equation}
The random observations $Z_t$ can be predicted by the random model response $Y_t$ besides a measurement error $E_t$. In this paper, the unknown model inputs $\vek{X}$ are material parameters which exhibit aleatory uncertainty as we are interested in the heterogeneous materials. The comparison of the stochastic and deterministic problems is depicted in Figure \ref{fig:det-stoch} together with associated uncertainties.

The paper complements the topic of identifying aleatory uncertainties discussed in the author’s habilitation thesis \cite{Kucerova:2020} and is organised as follows. Section \ref{sec:UQ} focuses on deterministic and probabilistic procedures available for identification of the heterogeneous material properties from indirect, noisy experimental data measured on an ensemble of specimens from the same material. Sections \ref{sec:Bayes} and \ref{sec:transformation} present two different formulations of stochastic inversion problem for aleatory uncertainties. The first formulation involves selecting a specific form of $f_{\vek{X}}(\cdot)$ for the unknown model parame\-ters, treating their statistical moments as uncertain random variables. The probabilistic identification is then focused on these so-called hyperparameters of the parameter distribution, which can be inferred by Bayesian approach~\cite{Gelman:2013}. The second formulation focuses especially on the joint probability distribution of the experimental data $f_{\vek{Z}}(\cdot)$. The distribution of the unknown parameters $f_{\vek{X}}(\cdot)$ is then defined by a~nonlinear transformation of $f_{\vek{Z}}(\cdot)$.

In this paper, the presented formulations of stochastic inversion are applied in two numerical examples. In Section \ref{sec:model1}, probabilistic parameter identification of the material model of kinematic and isotropic hardening for viscoplastic material is provided for synthetic data to verify the procedure, and then the model is calibrated from the real experimental data. This study case considers experimental data consisting of many highly correlated components. These are thus transformed first into uncorrelated principal components~\cite{Jolliffe:2002} with their joint distribution $f_{\vek{Z}}(\cdot)$ which is further transformed into $f_{\vek{X}}(\cdot)$. The second example elaborated in section \ref{sec:model2} shows a stochastic inversion solution for the case of processing observations from two different types of destructive experiments. Experi\-mental data obtained from different material samples do not allow the direct formulation of their joint distribution $f_{\vek{Z}}(\cdot)$ which complicates the parameter identification based on transformation. The section thus proposes a remedy by building $f_{\vek{Z}}(\cdot)$ using observed data in combination with modelled synthetic data. The concluding remarks are summarized in Section \ref{sec:conclusions}.

\section{Quantification of aleatory uncertainty}
\label{sec:UQ}
If no information is available about the morphology or grain size of the material constituents, but the goal is to model the material as a stochastic set (represen\-ting, for example, an ensemble of concrete specimens), the material parameters are no longer considered as deterministic. Identification of the material parameters leads to the quantification of aleatory uncertainty in the material parameters. The aleatory uncertainties are represented by random variables (or possibly random fields), whose probability distribution cha\-racterizes the distribution of values of the material properties within the ensemble.

Quantification of aleatory uncertainties in parameters of a nonlinear physical model is significantly less elaborated than quantification of epistemic uncertainty in deterministic parameters mentioned above. Documented strategies to the quantification of aleatory uncertainty in parameters of nonlinear models can be divided into two following categories. Even though it is possible to characterize the aleatory uncertainty in material parameters by random fields, most of the works describe each material property by a random variable. Each sample of material is modelled with one sample of the material property being spatially constant within the domain of a specimen. The distribution of material property in a set of material samples can be thus viewed as a distribution of some type of effective (or homogenized) property among the set of samples.

The first category of methods is based on determi\-nis\-tic fitting of aleatory uncertainty where the proba\-bility distribution prescribed to each material para\-me\-ter is modelled by a parametrized PDF $f_{\vek{X}}(\cdot)$. The goal of the inverse analysis then turns into a search for the parameters so-called hyperparameters of the prescribed PDF, which can be performed in a deterministic manner. \cite{Fonseca:2005:JSV} compares the mean-centered first-order perturbation method and Monte Carlo maximum likelihood estimation (MLE) applied to determine statistical moments of the prescribed PDFs. The proposed methods have several crucial limitations. First, the assumption of a known type of statistical model is very limiting and prohibits the identification of higher statistical moments. The perturbation method (even when extended by other nonlinear terms) is hardly applicable to nonlinear models and Monte Carlo-based MLE faces high computational requirements. Recently, \cite{Fang:2014:MSSP,Sepahvand:2014:PEM} employed independently polynomial chaos (PC)-based surrogates to accelerate the identification of aleatory uncertainties formulated as a deterministic optimization problem. In particular, statistical moments or PC coefficients defining the statistical model of physical parameters are optimized so as to fit the corresponding moments of model response to the moments obtained from the experiments. While \cite{Fang:2014:MSSP} emphasizes that the PC-based surrogate of a model response provides an efficient way for computing sensitivities, \cite{Sepahvand:2014:PEM} focuses on PC ability to effectively represent higher statistical moments and identify non-Gaussian parameters. Nevertheless, many related issues remain unsolved. For instance, quantification of related epistemic uncertainties is not considered at all. It means that the methodology is not able to reflect the number and the value of additional measurements. Also, the application of the proposed methods to the set of correlated observations, which are typically obtained by measu\-ring e.g. load-displacement curves (see Figure~\ref{fig:det-stoch}) or by collecting the observations from a set of probes placed on each specimen can be problematic. In such a situation, the higher dimensionality of the observations leads to the increasing complexity of their joint distributions and thus to an increasing dimensionality or a number of the underlying optimization problems.

On the other hand, probabilistic estimation of aleatory uncertainty enables to model aleatory uncertainty in material parameters together with related epistemic uncertainty in the correctness of estimated distributions. This approach is addressed in several recent works \cite{Arnst:2010:JCP,Mehrez:2012:MSSP2,Mehrez:2012:MSSP,Debruyne:2015:CS}. The authors use again the PC approximation of the distribution of material parameters, where the PC coefficients are obtained via the formulation of a likelihood function constructed for the measured data. Contrary to \cite{Fonseca:2005:JSV}, the PC coefficients are considered here as uncertain and identified in the Bayesian way by MCMC sampling. The introduction of new random variables for describing the PC coefficients thus allows us to quantify the epistemic uncertainty in the estimated distributions of aleatory material properties. Nevertheless, the method is derived only for a situation, where identified material parameters coincide with observed quantities. In the case of measuring a response of a nonlinear physical model, a set of deterministic inverse problems is solved first. Then quantification of aleatory uncertainty in estimated set of discrete parameter values follows in a separate step. Such separation is however not always feasible as the employed deterministic inverse problems are generally ill-posed. A similar concept was derived already much earlier for probabilistic estimation of statistical moments of PDFs prescribed to random variables  \cite{Gelman:2013,Sejnoha:2017:ES}. Nevertheless, its genera\-lization for estimation directly from observations of nonlinear response components was elaborated only recently in \cite{Behmanesh:2015:MSSP,Nagel:2016:PEM}, and its application to the estimation of viscoplastic material parameters was presented in \cite{Janouchova:2018:AP}. The principal limiting factor of this formulation for practical applications is the significant increase in problem dimensionality and computational effort with the number of material samples used to acquire the observations.

\section{Bayesian inference based on hierarchical modeling}
\label{sec:Bayes}
The core principle of Bayesian identification follows a~common reasoning pattern: the resulting belief about a random event is formed by combining all available information \cite{Bayes:1763}. This framework naturally incorporates uncertainty into our subjective knowledge of the identified parameters. Bayesian inference has become an increasingly popular and widely adopted approach to parameter identification, offering an ele\-gant solution to the inverse problem by making it well-posed.

The Bayesian inference can be extended to a multilevel framework, which also allows for the conside\-ration of aleatory uncertainties \cite{Nagel:2016:PEM,Behmanesh:2015:MSSP}. Hierarchical modelling enables to consider the model parameters $\vek{x}$ as stochastic variables, which for $i=1,...,n$ experi\-mental specimens have fixed yet unknown specific realizations $\vek{x}_i$. The structural prior knowledge about the material parameters is given as 
\begin{equation}
  (\vek{X}_i|\vek{\Theta}=\vek{\theta})\sim p_{\vek{X}|\vek{\Theta}}(\vek{x}_i|\vek{\theta}) \text{ for } i=1,...,n ,
\end{equation}
where the~probabilistic specification of the model parameters $\vek{x}$ is determined by new random variables, known as hyperparameters $\vek{\theta}$ \cite{Gelman:2013}, e.g. mean values and standard deviations. $\vek{\theta}$ are also fixed yet unknown and modelled as random variables with their own prior distribution called hyperprior
\begin{equation}
  \vek{\Theta}\sim p_{\vek{\Theta}}(\vek{\theta}).
\end{equation}

According to assumption of observation exchangeability \cite{Draper:1993}, the joint prior distribution is given as
\begin{equation}
  (\vek{X}_1,...,\vek{X}_n,\vek{\Theta})\sim \left(\prod_{i=1}^n{p_{\vek{X}|\vek{\Theta}}(\vek{x}_i|\vek{\theta}})\right) p_{\vek{\Theta}}(\vek{\theta}).
\end{equation}
The posterior distribution \textbf{$p(\vek{x}_1,\ldots,\vek{x}_n,\vek{\theta}|\vek{z}_1,\ldots,\vek{z}_n)$} is defined as
\begin{equation}
\left(\prod_{i=1}^n{f(\vek{z}_i|\vek{x}_i})\right)\left(\prod_{i=1}^n{p_{\vek{X}|\vek{\Theta}}(\vek{x}_i|\vek{\theta}})\right) p_{\vek{\Theta}}(\vek{\theta}) ,
\label{eq:posterior}
\end{equation}
where the likelihood function $f(\vek{z}_i|\vek{x}_i)$ expresses the difference between the measured values $\vek{z}_i$ and the model output $\vek{y}_i$ caused by independent measurement errors, here assumed to be normally distributed with zero mean values and experiment-specific covariance matrix $\vek{\sigma}_E\mat{I}$.
\begin{equation}
  (\vek{Z}_i|\vek{X}_i=\vek{x}_i)\sim f(\vek{z}_i-g(\vek{x}_i),\vek{\sigma}_E\mat{I}) \text{ for } i=1,...,n. 
\label{eq:likelihood}
\end{equation}
Our knowledge about the hyperparameters grows by updating based on every new measurement while structure of the model parameters' prior distribution remains unchanged.

\begin{figure*}[b]
\centering
\includegraphics[width=0.9\linewidth]{ 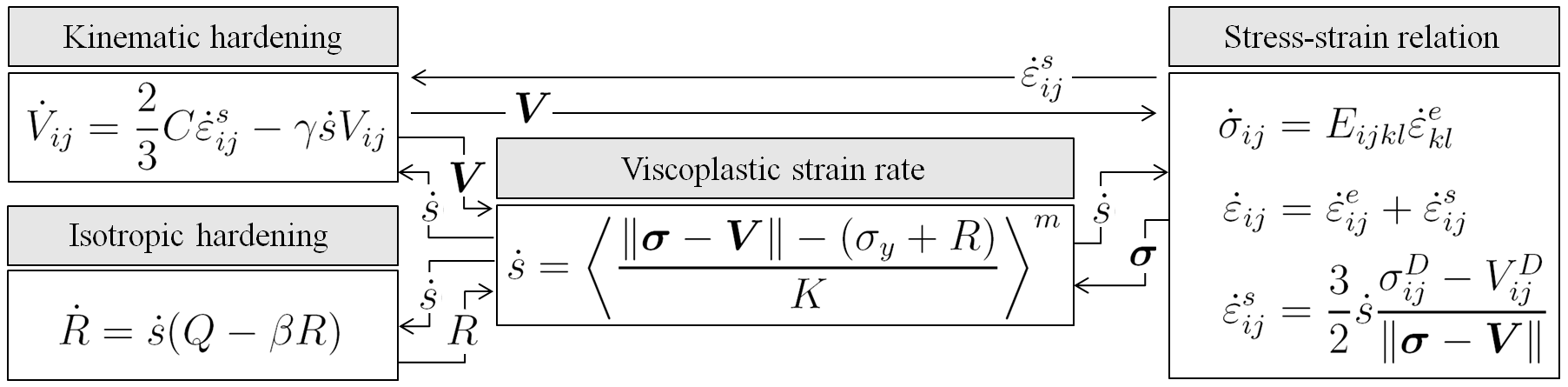} 
\caption{Constitutive equations of the viscoplastic model with non-linear isotropic and kinematic hardening.}
\label{fig:model}
\end{figure*}

\section{Nonlinear transformation of random variables} 
\label{sec:transformation}
A key drawback of the aforementioned formulation for quantifying aleatory uncertainty lies in the requirement to assign a specific distribution family to each material parameter and to determine only its statistical moments. This can be partially relaxed by describing each random material parameter by polynomial function and searching for its coefficients, thus allowing one to find some more general probability distribution. The preliminary choice of the particular polynomial degree with the given number of polynomial coefficients is, in fact, very similar limitation to the choice of a distribution family with some given number of statistical moments. Driven by this motivation, another computationally more effective strategy is presented in the following text. Its simple idea is based on nonlinear transformation of random variables also known as change of variables; see \cite{Marcy:2022} for more detailed theoretical overview.

In this paper, we consider the numerical model (\ref{eq:model}) with nonlinear mapping from parameter space to response space 
\begin{equation}
    \bf{g}:(\mathcal{X}\subseteq \mathbb{R}^{n_x}) \mapsto (\mathcal{Y}\subseteq \mathbb{R}^{n_z}),
    \label{eq:mapping}
\end{equation}
where the model response $\vek{y}$ represents a $n_z$-dimen\-sional vector of a prediction of observable data assumed to be a random vector with a joint probability distribution $\vek{Z}\sim f_{\vek{Z}}(\vek{z})$. The $n_x$-dimensional vector of model parameters is considered as a random vector with a joint probability distribution to be determined $\vek{X}\sim f_{\vek{X}}(\vek{x})$. The solution of stochastic inversion can be obtained by a nonlinear transformation of probability distribution $f_{\vek{Z}}({\vek{z}})$ into distribution $f_{\vek{X}}({\vek{x}})$ according to the following relation
\begin{equation}
    f_{\vek{X}}(\vek{x}) = f_{\vek{Z}}({\bf g}(\vek{x}))\cdot|J_{\vek{y}} (\vek{x})|,
    \label{eq:transform}
\end{equation}
where $J_{\vek{y}} ({\vek{x}})$ denotes the Jacobian $\frac{\partial \bf g(\vek{x})}{\partial \vek{x}}$. Making this change of variables explicit imposes certain constraints on the model ${\bf g}({\vek{x}})$. First, the inverse image support $\mathcal{X}$ of the distribution $f({\vek{x}})$ must be contained within the range of ${\vek{y}}$, i.e. $\mathcal{Y} = {\bf g}^{-1}({\bf x})$. Second, ${\bf g}({\vek{x}})$ must be a differentiable transformation from $\mathcal{X}$ to $\mathcal{Y}$ with a~differentiable inverse. The constraints do not be fully fulfilled by the original material model, but they can be easily valid for its polynomial approximation \cite{Janouchova:2018:AM} employed for the acceleration of the transformation process. In such a case, the Jacobian can be easily obtained analytically. Moreover, the second constraint is not a great liability, when the transformation is performed by means of suitable sampling methods such as Markov chain Monte Carlo, see \cite{Marzouk:2007:JCP}.

The crucial condition of this formulation is that the mapping (\ref{eq:mapping}) must be bijective. In practice, laboratory experiments often \R{result} in load-deflection curves discretised into a set of points corres\-ponding to observed vector $\vek{z}$. The dimension $n_z$ is thus typically much higher than the dimension $n_x$ and particular components of ${\vek{z}}$ are highly correlated, see Figure \ref{fig:det-stoch}b. Once the data from $n$ repetitions of the experiments are collected into the matrix ${\bf Z} = ({\vek{z}}_1, \dots, {\vek{z}}_{n}) \in \mathbb{R}^{n_z \times n}$, the correlations between couples of particular response components can be computed directly from the data. By means of principal component analysis (PCA) they can be easily transformed into a set of uncorrelated quantities and ordered according to their particular variance \cite{Jolliffe:2002}. The first few components account for most of the statistical variability in all of the original
data. PCA thus allows to reduce the dimensionality of observed data ${\bf Z}$ to the set ${\bf Q}  = ({\vek{q}}_1, \dots, {\vek{q}}_{n}) \in \mathbb{R}^{n_x \times n}$ of first $n_x$ most important components without any significant information loss. 

\R{The number of retained principal components is chosen to match the dimension of the parameter vector in order to preserve a well-conditioned and locally bijective mapping between the reduced data representation and the parameter space. Components associated with small eigenvalues correspond to directions of low variance in the response space and are typically dominated by noise or weakly identifiable features. Discarding them therefore acts as a regularization step and does not remove information that could be reliably exploited by a model with the given number of parameters.} 

\R{In the present formulation, measurement error is not modeled explicitly and is absorbed into the variability of the data. The reduction to the first $n_x$ principal components intentionally discards the remaining variability, including higher-order response modes and part of the noise. The resulting parameter uncertainty should therefore be interpreted as an effective uncertainty.} The solution for a case $n_x < n_z$ is following
\begin{equation}
    f_{\vek{X}}({\vek{x}}) = f_{\vek{Q}}(\vek{q}(\bf g(\vek{x}),\mathbf{Z}))\cdot|\frac{\partial {\vek{q}(\bf g(\vek{x}),\mathbf{Z})}}{\partial \vek{x}}|
    \label{eq:transform_PCA}
\end{equation}
The uncorrelation of the principal components, however, do not imply their statistical independence. Therefore, the construction of their joint probabi\-lity density function $f_{\vek{Q}}({\bf q})$ is again not trivial and one has to keep in mind that its decomposition into a product of independent marginal distributions $f_{\vek{Q}}({\bf q}) = \prod f_{\vek{Q}_i}(q_i)$ may contain certain error. 

\section{Estimation of parameter distributions from cyclic loading tests}
\label{sec:model1}

The presented formulations of the stochastic inversion problem are employed in the calibration of a numerical model of kinematic and isotropic hardening for a~copper alloy, as defined by the constitutive equations shown in Figure~\ref{fig:model}.

\begin{figure}[h!]
\centering
\begin{tabular}{cc}
\includegraphics[width=0.9\linewidth]{ 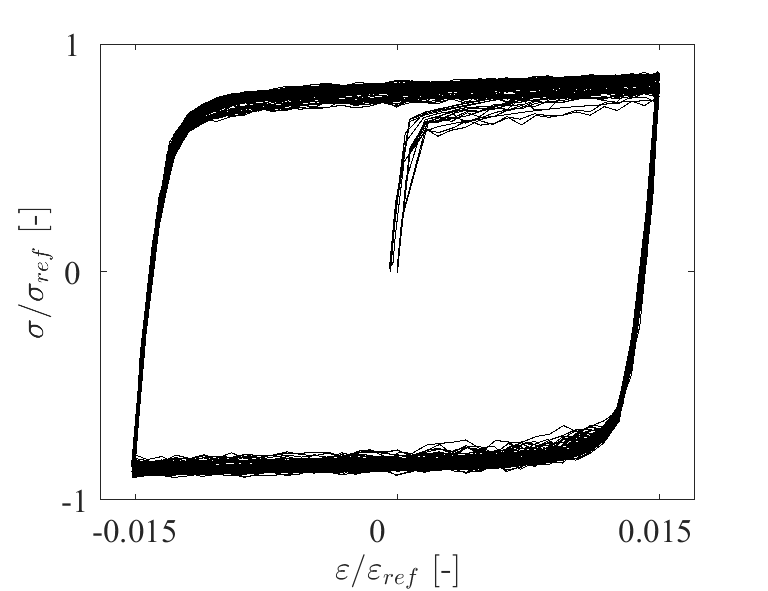} \\
\includegraphics[width=0.9\linewidth]{ 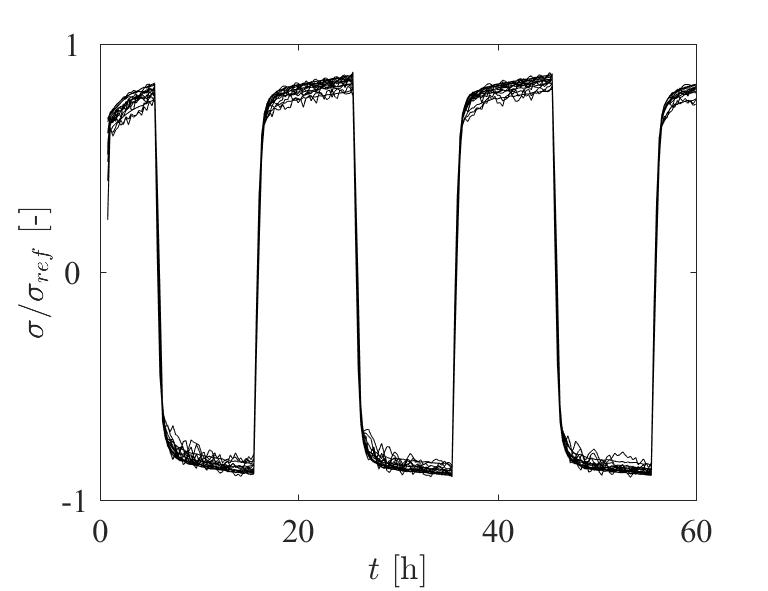} \\
\end{tabular}
\caption{Experimental data: stress-strain curves.}
\label{fig:exper}
\end{figure}

The available experimental data, corresponding to sixteen repetitions of a strain-controlled cyclic loading test at a high temperature, are shown in Figure~\ref{fig:exper}.

The material data are expressed in relative units: stresses are given as ratios to $\sigma_{ref}$ and strains as ratios to $\varepsilon_{ref}$. These reference values remain constant throughout this paper, ensuring consistency across all figures. The example presented here was part of a~project conducted under the European Space Agency, where the results are confidential; therefore, all data have been scaled. The stress-strain curves are discretised into 568 points, and the experimental errors at each measured point are assumed to be zero-mean i.i.d.\ Gaussian random variables.
At higher temperatures, the cyclic material behaviour is dominated by the kinematic hardening mechanism, with stress saturation reached after only a few cycles~\cite{Schwarz:2011}. The copper alloy does not exhibit a distinct yield limit, and its kinematic hardening response is non-linear.

The corresponding model has six material parameters and the model itself is considered as a black box in the identification procedures. The hardening moduli $Q$ and $C$, given in equations in Figure~\ref{fig:model}, is replaced by the saturation values $R_{\infty}={Q}/{\beta}$ and $V_{\infty}={C}/{\gamma}$. The model parameters 
\begin{equation}
  \vek{X} = (..., X_j,...)=(\sigma_y, V_{\infty}, \gamma, R_{\infty}, \beta, \nu)
\label{eq:model_par}
\end{equation}
to be identified are summarized in Table \ref{tab:par} and their feasible domain, where the model is stable and produces valid predictions, is scaled into $[0;1]$. In order to model behaviour of a heterogeneous material, the parameters are stochastic and the calibration procedure focuses on identifying their probability distribution expressing the corresponding aleatory uncertainty.
\begin{table}[h!]
\centering
\tabcolsep=2pt
\begin{tabular}{cl}
\toprule 
\bfseries Parameter & \bfseries Name \\
\midrule
\R{$\sigma_y$}  & Yield stress \\
\R{$V_{\infty}$} & Kinematic hardening saturation \\
\R{$\gamma$} & Kinematic dynamic recovery constant \\
\R{$R_{\infty}$}  & Isotropic hardening saturation \\
\R{$\beta$} & Isotropic dynamic recovery constant \\
\R{$\nu$} & Poisson ratio \\\bottomrule
\end{tabular}
\caption{Model parameters to be identified.}
\label{tab:par}
\end{table}

\subsection{Verification of identification procedures}
\label{verification}

\begin{figure}[h!]
\centering
\begin{tabular}{c}
\includegraphics[width=0.9\linewidth]{ 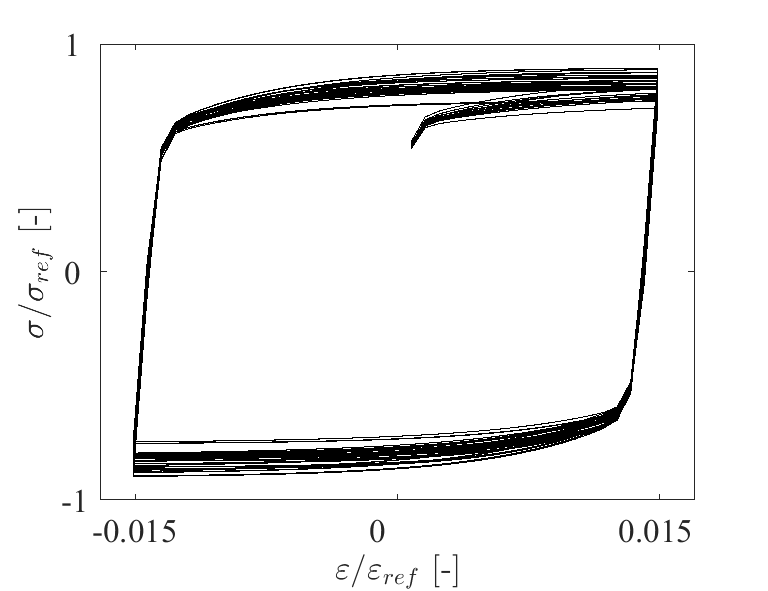} \\
\includegraphics[width=0.9\linewidth]{ 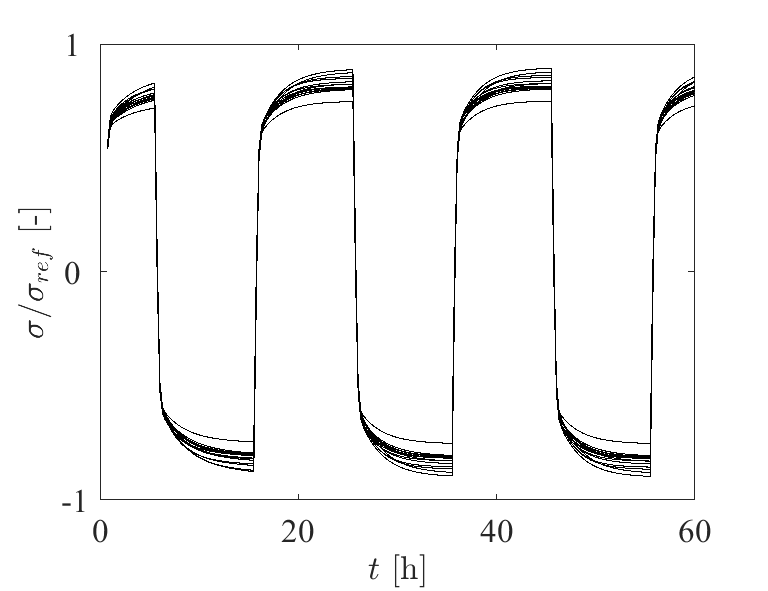} \\
\end{tabular}
\caption{Sixteen synthetic curves obtained from the numerical model of the cyclic loading test.} 
\label{fig:synthetic_data}
\end{figure}

An ensemble of $16$ curves was simulated to represent synthetic data from the cyclic loading test on a hete\-rogeneous viscoplastic material; see Figure~\ref{fig:synthetic_data}. This data serves as a testing set for the verification of both identification formulations.

The ensemble of synthetic data is produced for independent $n = 16$ samples of inputs $\vek{x_i}$, $i=1,...,n$, from the lognormal distribution with prescribed mean values and standard deviations. One parameter is insignificant in the elaborated experiment and does not influence the model response at all; its value is intentionally fixed in order to demonstrate how the identification procedures cope with such a circumstance. The aim is to identify the relevant parameters in the given loading test and to determine their probability distributions corresponding to that of the synthetic data.

\subsubsection{Bayesian hierarchical model}

\begin{figure*}[b]
\centering
\includegraphics[width=1.0\linewidth]{ 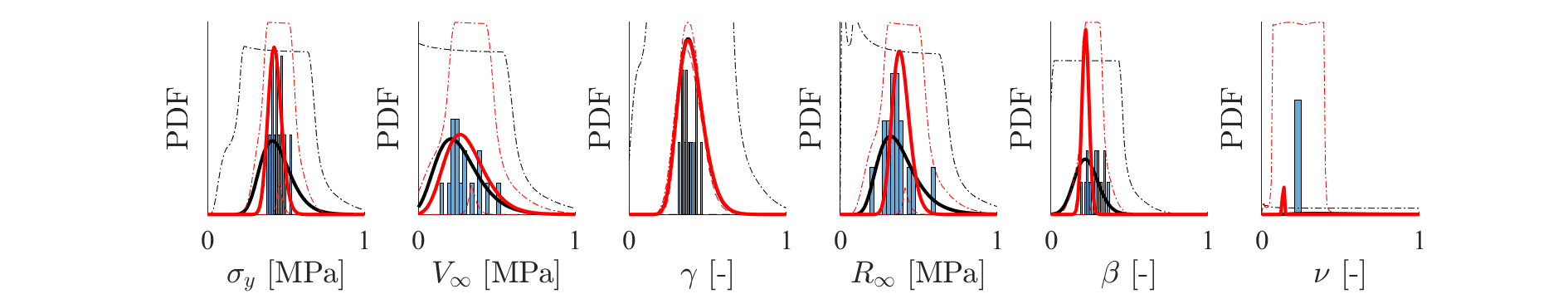} 
\caption{Hierarchical Bayes - verification: Histograms of synthetic experimental inputs, prior (black) and posterior (red) parameters' PDFs corresponding to hyperparameters' mean of prior and mode of posterior (solid lines) accompanied by bounds corresponding to $90$ \% of  hyperparameters' values (dashed lines).}
\label{fig:verifikace_parametry}
\end{figure*}
\begin{table*}[b]
\centering
\tabcolsep=2pt
\begin{tabular}{cc|c|c|c|c|c|c|c|c|c|c|c|c|}
& & \multicolumn{2}{c|}{$\sigma_y$ [MPa]} & \multicolumn{2}{c|}{$V_\infty$ [MPa]} & \multicolumn{2}{c|}{$\gamma$ [-]} & \multicolumn{2}{c|}{$R_\infty$ [MPa]} & \multicolumn{2}{c|}{$\beta$ [-]} & \multicolumn{2}{c|}{$\nu$ [-]}\\
& & $\mu$ & $\sigma$ & $\mu$ & $\sigma$ & $\mu$ & $\sigma$ & $\mu$ & $\sigma$ & $\mu$ & $\sigma$ & $\mu$ & $\sigma$\\
\hline
\multirow{3}{*}{MEAN} 
 & PRESCRIBED & $0.444$ & & $0.263$ & & $0.386$ & & $0.381$ & & $0.283$ & & $0.240$ &\\
 & DATA & $0.444$ & & $0.292$ & & $0.378$ & & $0.361$ & & $0.282$ & & $0.240$ &\\
 & PRIOR & $0.444$ & $0.128$ & $0.263$ & $0.182$ & $0.397$ & $0.146$ & $0.381$ & $0.165$ & $0.228$ & $0.132$ & $0.520$ & $0.300$\\
 & POSTERIOR & $0.457$ & $0.025$ & $0.321$ & $0.039$ & $0.394$ & $0.002$ & $0.404$ & $0.032$ & $0.268$ & $0.016$ & $0.172$ & $0.049$\\ 
 \hline
 \multirow{3}{*}{STD} 
 & PRESCRIBED & $0.052$ & & $0.105$ & & $0.048$ & & $0.095$ & & $0.044$ & & $0$ &\\
 & DATA & $0.040$ & & $0.098$ & & $0.041$ & & $0.090$ & & $0.049$ & & $0$ &\\
 & PRIOR & $0.102$ & $0.037$ & $0.145$ & $0.053$ & $0.082$ & $0.046$ & $0.131$ & $0.048$ & $0.076$ & $0.031$ & $0.300$ & $0.173$\\
 & POSTERIOR & $0.043$ & $0.006$ & $0.069$ & $0.021$ & $0.083$ & $0.003$ & $0.068$ & $0.019$ & $0.028$ & $0.007$ & $0.110$ & $0.120$\\ 
 \hline
\end{tabular}
\caption{Prior and posterior mean values and standard deviations of hyperparameters in comparison with prescribed values for generating the synthetic data and true statistical moments of the data.}
\label{tab:Ver_HB_hyp}
\end{table*}

Bayesian identification requires an assumption about the parameter distribution. In this case, the unknown model parameters are assumed as lognormally distributed random variables
\begin{equation}
\begin{aligned}
  (X_j|\vek{\Theta}_j=(\mu_{X_j}, \sigma_{X_j}))\sim LN(X_j; \mu_{X_j}, \sigma_{X_j}^2) \\
  \text{ for } j=1,...,6 , \\
\end{aligned}
\end{equation}
and the corresponding mean value ${\mu}_{{X}_j}$ and standard deviation $\vek{\sigma}_{{X}_j}$ are the unknown hyperparameters $\vek{\Theta}_j=(..,\Theta_{jh},...)$ with uninformative uniform hyperprior $U(\Theta_{jh},a_{jh},b_{jh})$.
Here, the model parameters are considered to be uncorrelated, otherwise another hyperparameters have to be introduced; see e.g. \cite{Janda:2024}. The joint prior distribution is 
\begin{equation}
  \left(\prod_{i=1}^{n}\prod_{j=1}^{n_X}LN(x_{ij};\vek{\theta_j})\right) \prod_{j=1}^{n_X}\prod_{h=1}^{n_H}{U(\theta_{jh};a_{jh},b_{jh})}.
\label{eq:prior_example}
\end{equation}

Likelihood function (\ref{eq:likelihood}) involves all the synthetic data ($n=16$, $n_t=568$) and corresponding i.i.d. Gaussian experimental errors with zero mean values and identic standard deviation $\sigma_E = 0.01$ at every point of the curves%
\begin{equation}
  \prod_{i=1}^{n}{f(\vek{z}_i|\vek{x}_i}) = \prod_{i=1}^{n}{N(\vek{z}_i-g(\vek{x}_i); \vek{0}, \vek{\sigma}_E\mat{I})}. 
\label{eq:likelihood_example}
\end{equation}
\R{The value of $\sigma_E$ is assumed to be known and fixed, as it is specified by the provider of the experimental data and the same noise level is also used for the pseudo-experimental data.}

(\ref{eq:prior_example}) and (\ref{eq:likelihood_example}) are combined according to Bayes' rule into the joint posterior distribution (\ref{eq:posterior}) of $108$ random variables, specifically, $96$ parameters ($n_X=6$ model parameters corresponding to each of $16$ measurements) and $12$ hyperparameters ($n_H=2$ statistical moments of every single parameter's marginal PDF)

The Bayesian inference is realized by the Markov chain Monte Carlo (MCMC) method providing posterior samples~\cite{Spall:2003, Marzouk:2007:JCP} by Metropolis algorithm with a~symmetric proposal distribution. \R{The posterior samples were obtained from a single Markov chain after discarding an initial burn-in period. The proposal variance in the Metropolis algorithm was tuned through repeated preliminary runs to achieve an acceptance rate between approximately $20$ \% and $50$ \%, ensuring satisfactory mixing of the chain \cite{Rosenthal:2011}.} The computational effort is reduced by substituting the full numerical model by its polynomial chaos-based approximation, specifically by the Hermite polynomials of standard normal random variables~\cite{Wiener:1938, Xiu:2002:SIAM}. The polynomial coefficients are computed with a help of regression method \cite{Blatman:2010:PEM} from $10000$ model simulations based on Latin Hypercube sampling~\cite{Iman:1980} from prior parameter distribution. On the basis of cross-validation results, the eighth degree of Hermite polynomials is used.

The identified marginal PDFs of each parameter are depicted in Figure \ref{fig:verifikace_parametry} compared to the histograms of the synthetic data inputs generated from the prescribed lognormal distribution. \R{The displayed marginal densities are not prior- or posterior-predictive distributions. Instead, they represent the aleatory uncertainty corresponding to a representative choice of hyperparameters, while the epistemic uncertainty is reflected by the spread of the hyperparameters' distribution. For the prior depicted in black, whose hyperparameters are assumed uniformly distributed over the feasible domain, the mean value is used as a central illustrative reference.}
The red solid lines correspond to parameters' marginal PDFs with identified maximum a posteriori (MAP) estimates of hyperparameters \R{and represent the most probable identified distribution of the aleatory uncertainties in the parameters' values. The dashed} lines form an envelope of parameters' PDFs corresponding to $90$ \% of  hyperparameters' values representing the epistemic uncertainties. The apparent distance of dashed lines depicts a significant epistemic uncertainty about the identified PDF as it is in the case of parameters $V_{\infty}$, $R_{\infty}$ and $\nu$, while the PDF of $\gamma$ is identified very certainly and the dashed lines are very close to the solid line.

\begin{figure*}[b]
\centering
\includegraphics[width=0.8\linewidth]{ 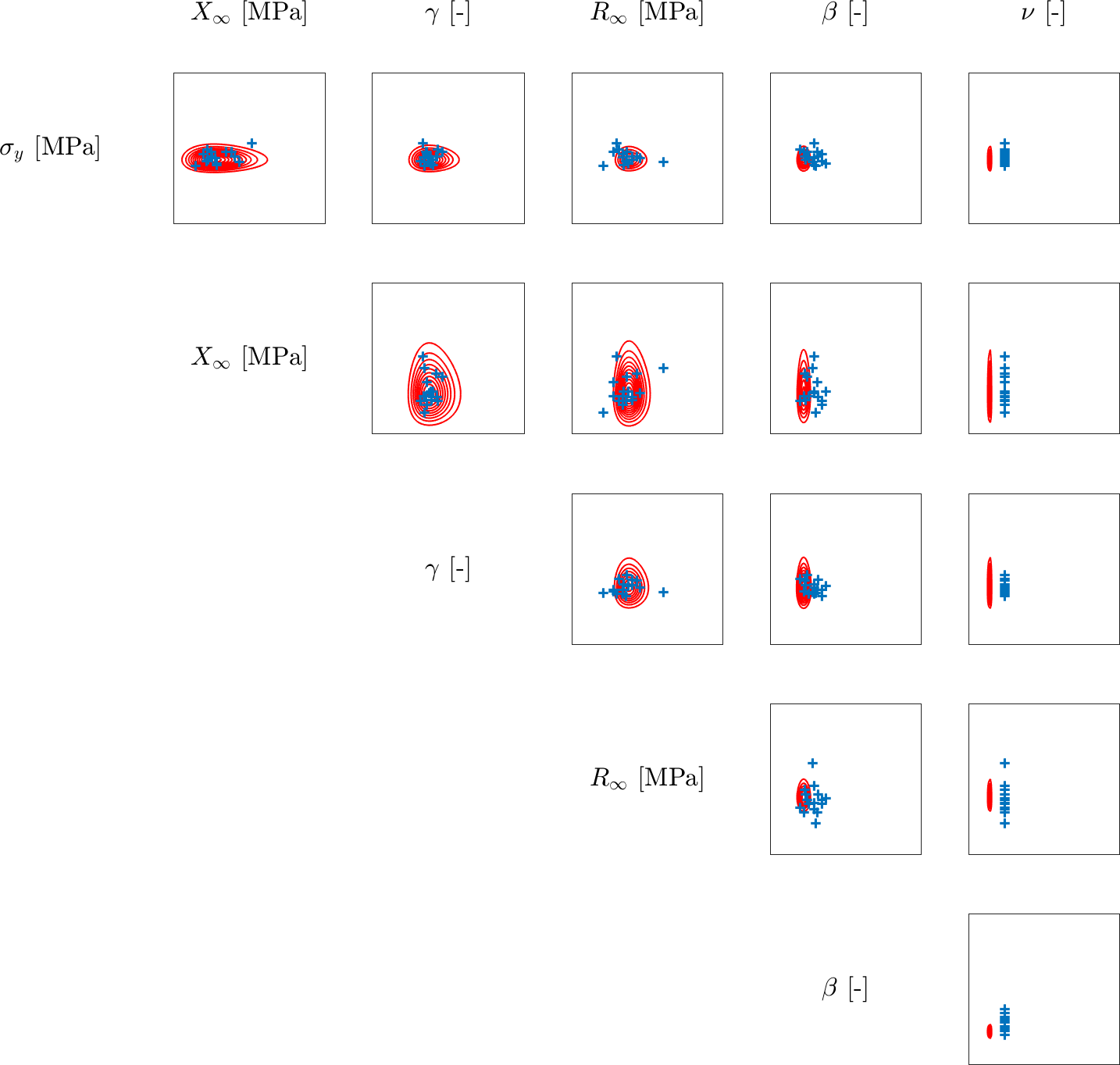} 
\addtocounter{figure}{1}
\caption{Hierarchical Bayes - verification: Identified parameters' 2D marginal PDFs corresponding to MAP estimation of hyperparameters and synthetic experimental inputs.}
\addtocounter{figure}{-1}
\label{fig:verifikace_parametry_2d}
\end{figure*}

Table \ref{tab:Ver_HB_hyp} summarizes numerical results for the hyperparameters (MEAN, STD). The prescribed values used for generating the synthetic data are supplemented by the true values calculated from the obtained data ensemble, as the relatively small data sample size may render the prescribed values potentially misleading. To assess the identification of aleatory uncertainty, the mean (\text{MEAN}) and standard deviation (\text{STD}) of the data are compared with the posterior mean values \( \mu \).  
The best agreement for both statistical moments is achieved in the case of \( \sigma_y \).  
Moreover, the mean values for the parameters \( \gamma \) and \( \beta \) are identified with very small errors—within \( 2\% \) of the considered interval.  
A~reduction in epistemic uncertainty due to the identification process is evident from the comparison of the prior and posterior standard deviations \( \sigma \).  
As noted above, the lowest level of epistemic uncertainty is observed for the kinematic hardening parameter \( \gamma \), which aligns with expectations from the experimental setup, where cyclic material behaviour at high temperatures is dominated by the kinematic hardening mechanism.  
The most uncertain posterior PDF corresponds to the parameter \( \nu \), whose statistical moments are also associated with the largest identification errors.  
The higher level of epistemic uncertainty associated with \( \nu \) suggests its limited importance for the model response; however, its insignificance can be confirmed more robustly via global sensitivity analysis \cite{Saltelli:2000}, see Figure~\ref{fig:SA_SRCC}.  
The sensitivities, expressed by the Spearman rank correlation coefficient (SRCC), are computed using the same prior data set that was used for constructing the surrogate model, incurring no additional computational cost—yet the insight provided by this analysis is considerable.  
A zero SRCC value between parameter \( \nu \) and each component of the model response confirms that \( \nu \) is an insignificant parameter. \R{Since parameter $\nu$ does not influence the model response, it is structurally non-identifiable in the considered experiment and its posterior distribution should theoretically coincide with the prior. Any slight concentration observed in the inferred posterior can be attributed to numerical effects associated with the surrogate approximation and finite MCMC sampling rather than to genuine information contained in the data.}
\begin{figure}[h!]
\centering
\includegraphics[width=0.9\linewidth]{ 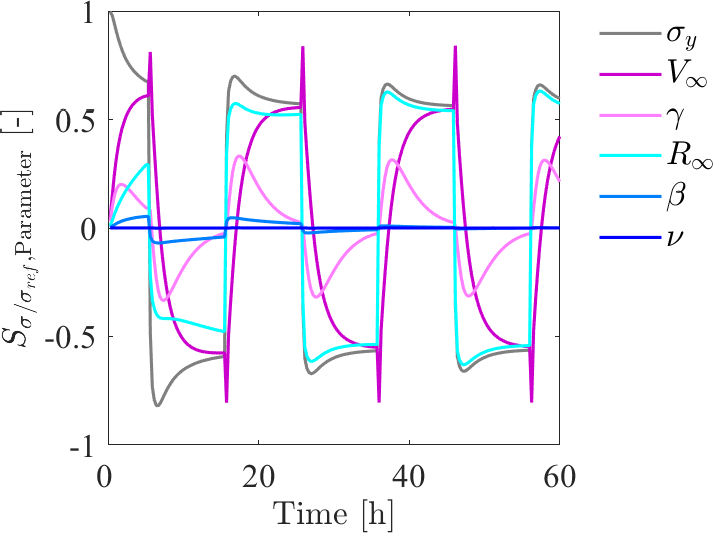} 
\addtocounter{figure}{-1}
\caption{Global sensitivity analysis based on Spearman rank correlation coefficient.}
\addtocounter{figure}{1}
\label{fig:SA_SRCC}
\end{figure}

Figure \ref{fig:verifikace_parametry_2d} presents the marginal PDFs of the parameters in two dimensions, obtained using the MAP estimate. Under the prescribed assumptions, the parameters are mutually independent as well as it is prescribed for generating the synthetic data. The identified aleatory uncertainty appears to be underestimated for $V_{\infty}$, $R_{\infty}$ and $\beta$. Although the parameter $\nu$ is not clearly identifiable, its estimated PDF is still valid, just like any other PDF associated with this parameter.

\begin{figure}[h!]
\centering
\includegraphics[width=1\linewidth]{ 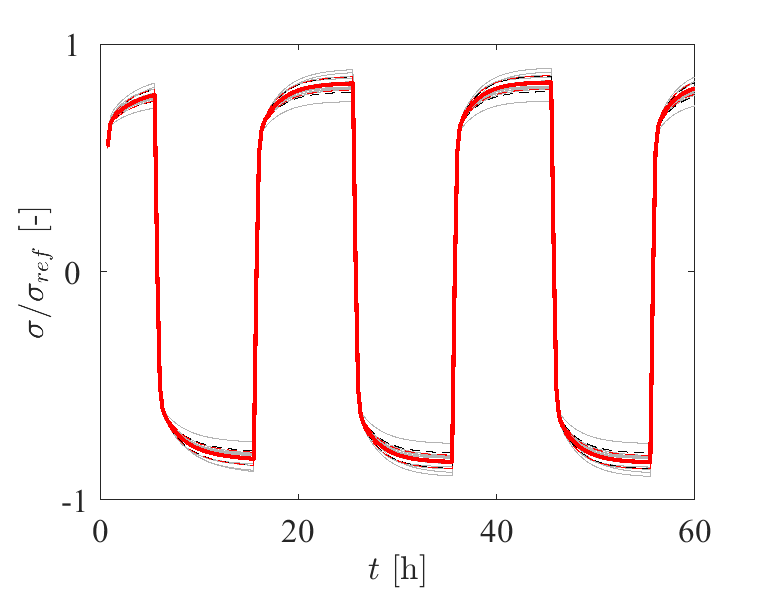} \\
\caption{Hierarchical Bayes - verification: Comparison of synthetic observations (black) with model responses (red) corresponding to the identified MAP parameters' distribution in a form of mean value $\pm$ standard deviation.}
\label{fig:verifikace_napeti}
\end{figure}

\begin{figure}[h!]
\centering
\includegraphics[width=1\linewidth]{ 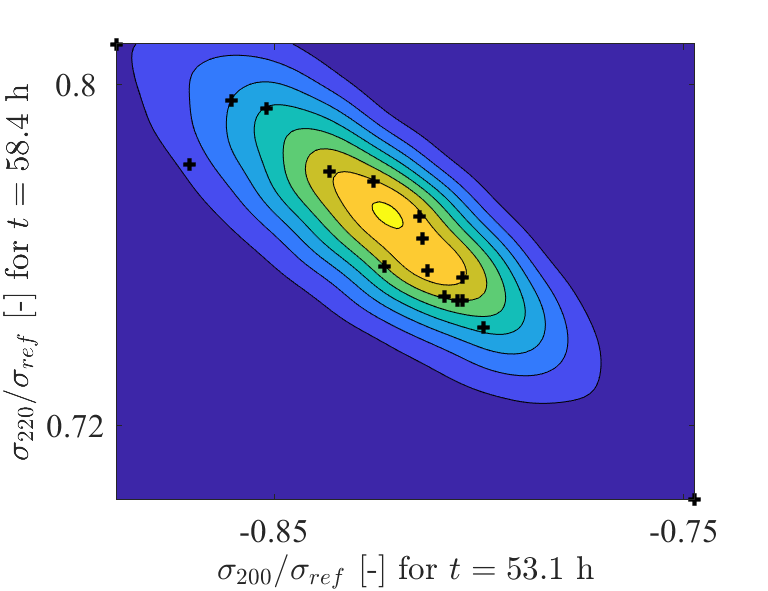} 
\caption{Hierarchical Bayes - verification: Comparison of synthetic observations (black) with isolines of joint PDF of model responses for $t = \R{53.1}$ h and $t = 58.4$ h corresponding to the MAP parameters' distribution.}
\label{fig:verifikace_napeti2}
\end{figure}

The identified parameters' PDF corresponding to hyperparameters' MAP values is used to quantify aleatory uncertainty in the model response, which is presented in Figure \ref{fig:verifikace_napeti}. 
Estimation of the aleatory uncertainty in the model response has a good match with the synthetic experimental data in their mean values and standard deviations of particular response components, see Figure \ref{fig:verifikace_napeti}. However, the response components exhibit a complex joint probability distribution. For the sake of clarity, Figure \ref{fig:verifikace_napeti2} 
shows the two-dimensional marginal distribution of two selected response components in comparison with synthetic observations, where agreement in correlation is achieved. From this point of view, the results of the proposed identification procedure are satisfactory. 

\subsubsection{Transformation of variables}

The identification of material heterogeneity is now performed for the same data set by the second formulation based on transformation of variables. In this case, the initial information about the identified parameters reflects their feasible domain defined by the feasible intervals scaled into $[0;1]$. In contrast to Bayesian formulation, there are no additional assumptions about unknown parameters' PDF, but the identification is based on the prescription of PDF to the data. 

The data obtained by discretization of the curves are first transformed into six of the most important principal components because the six model parameters are subject to identification. The marginal distribution of each principal component is formulated as normal distribution with mean and standard deviation estimated from the samples and their joint distribution is obtained as a product of marginals under the simplifying assumption of their mutual independence. Joint distribution of material parameters is then obtained by MCMC sampling of the (\ref{eq:transform}).

\begin{figure*}
\centering
\includegraphics[width=1.0\linewidth]{ 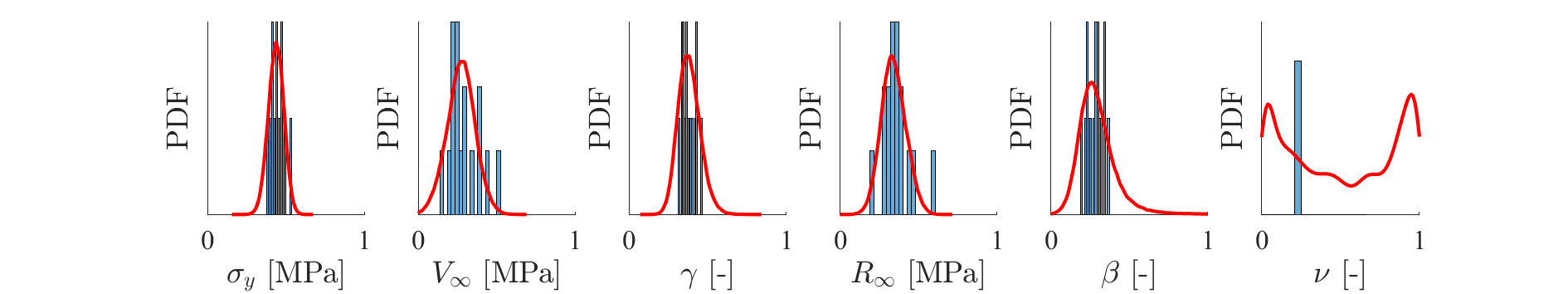} 
\caption{Transformation - verification: Histograms of synthetic experimental inputs and identified (red) parameters' marginal univariate PDFs.}
\label{fig:verifikace_parametry_PCA}
\end{figure*}
\begin{figure*}
\centering
\includegraphics[width=0.8\linewidth]{ 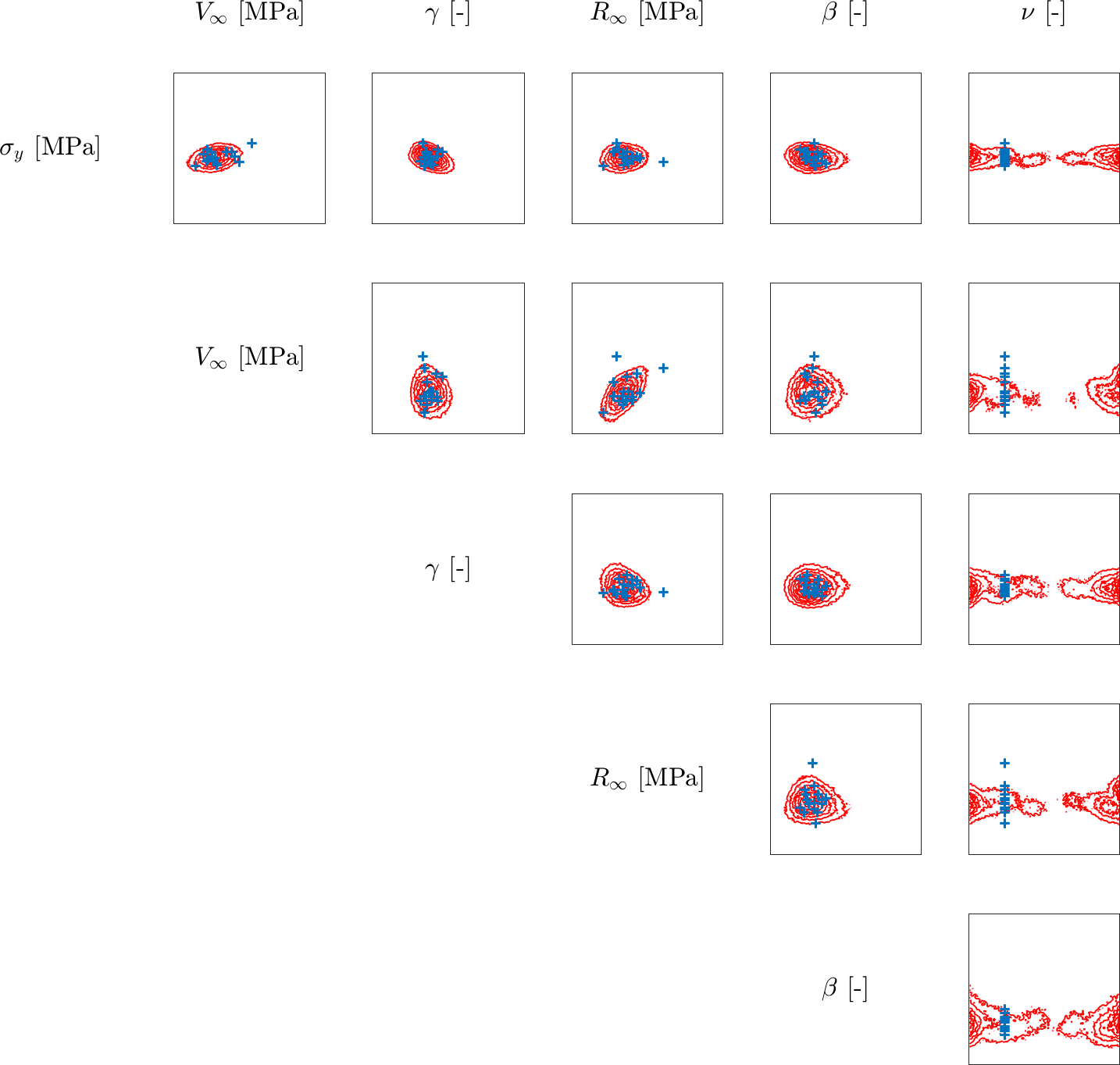} 
\caption{Transformation - verification: Identified parameters' 2D marginal PDFs corresponding to MAP estimation of hyperparameters and synthetic experimental inputs.}
\label{fig:verifikace_parametry_2d_PCA}
\end{figure*}

The obtained probability density functions of material parameters are shown in Figure~\ref{fig:verifikace_parametry_PCA}. The results confirm that the proposed identification procedure is able to determine the true distributions of five relevant material parameters precisely enough, including correlation between the parameters $V_{\infty}$ and $R_{\infty}$, see Figure~\ref{fig:verifikace_parametry_2d_PCA} presenting identified two-dimensional parameters' PDFs. Although the synthetic inputs are generated from prescribed independent distributions, a small data set leads to the correlated samples, the obtained correlation between $V_{\infty}$ and $R_{\infty}$ is $0.4$, the identified one is $0.6$. A different distribution is obtained only for sixth parameter, which is insignificant in the simulated experiment and the data contain no information about it, thus any distribution of $\nu$ is valid to produce the correct distribution of the model response. 

\begin{figure}[h!]
\begin{center}
\includegraphics[width=1\linewidth]{ 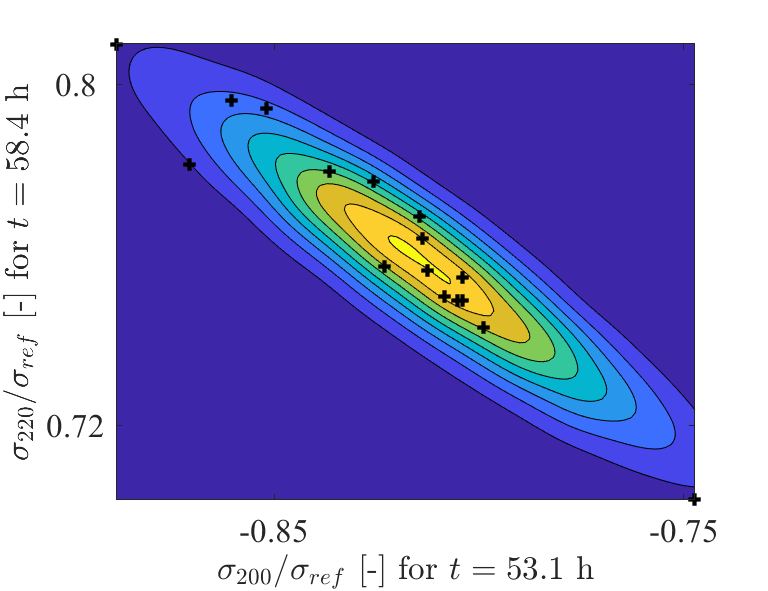}
\end{center}
\caption{Transformation - verification: Comparison of synthetic observations (black) with isolines of formulated joint PDF of model responses for $t = \R{53.1}$~h and $t = 58.4$ h.} 
\label{fig:verifikace_odezva_PCA2}
\end{figure}

The model predictions for the identified distribution of random material inputs are compared with the prescribed synthetic data in 
\begin{figure}[h!]
\begin{center}
\includegraphics[width=1\linewidth]{ 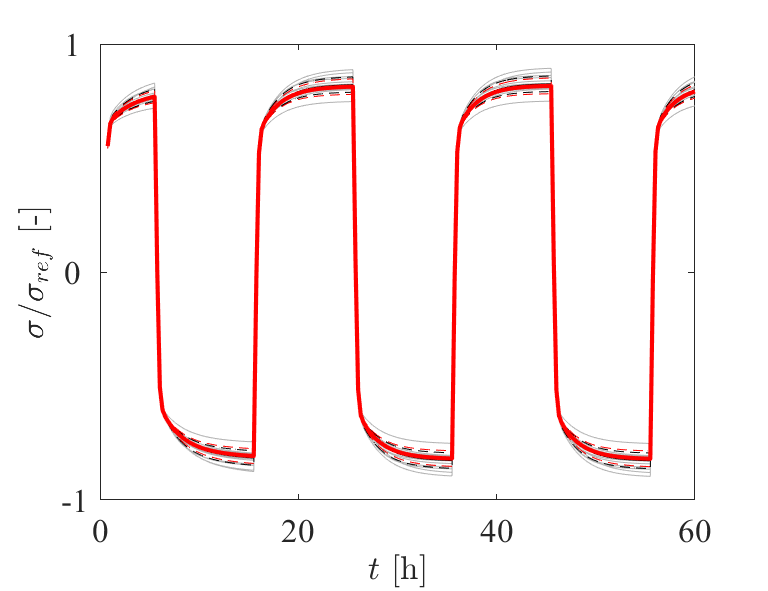}
\end{center}
\caption{Transformation - verification: Comparison of synthetic observations (black) with model res\-ponses (red) corresponding to the identified parameters' distribution in a form of mean value $\pm$ standard deviation.} 
\label{fig:verifikace_odezva_PCA}
\end{figure}
Figures~\ref{fig:verifikace_odezva_PCA} and \ref{fig:verifikace_odezva_PCA2}, where a high level of agreement is achieved.

\subsubsection{Comparison}
To summarize the results obtained from both identification procedures, the representative numerical values are listed in Tables \ref{tab:Ver_comparison_sigma} and \ref{tab:Ver_comparison_errors}. Table \ref{tab:Ver_comparison_sigma} presents the statistical moments (mean values (MEAN), standard deviations (STD) and correlations (CORR)) of the chosen response components $\sigma_{200}$ and $\sigma_{220}$ calculated on the basis of $10000$ full numerical model simulations for inputs $\vek{X}$ from their identified PDF. 
\begin{table}[!h]
\centering
\tabcolsep=3pt
\begin{tabular}{c|rr|rr|r|}
& \multicolumn{2}{c|}{MEAN} & \multicolumn{2}{c|}{STD} & CORR\\
& $\sigma_{200}$ & $\sigma_{220}$ & $\sigma_{200}$ & $\sigma_{220}$ & $\sigma_{200}, \sigma_{220}$\\
 \hline
 Data & $-0.823$ & $0.765$ & $0.034$ & $0.026$& $-0.954$\\
 Bayes & $-0.828$ & $0.772$ & $0.029$ & $0.023$ & $-0.825$\\ 
 Transf.& $-0.813$ & $0.757$ & $0.033$ & $0.024$ & $-0.952$ \\
 \hline
\end{tabular}
\caption{Comparison of statistical moments of the observed data and the predictive model responses for both identification formulation.}
\label{tab:Ver_comparison_sigma}
\end{table}

The verification results demonstrate strong agreement between the data and the model responses for both approaches. Specifically, Bayesian identification provides more accurate mean values of the predictions, while the standard deviations and correlation reach a better match for transformation-based identification. This trend is also confirmed in Table \ref{tab:Ver_comparison_errors} by values of mean absolute errors in predicted mean (MEAN) of each response component $\overline{y_t}$ and its value in the observed data $\overline{z_t}$
\begin{equation}
  \text{MAE} = 1/568 \sum_{t=1}^{568}{|\overline{z_t}-\overline{y_t}|}
\label{eq:MAE}
\end{equation}
and mean absolute percentage errors 
\begin{equation}
  \text{MAPE} = \frac{100\%}{568} \sum_{i=1}^{568} \left| \frac{\overline{z_t} - \overline{y_t}}{\overline{z_t}} \right|
\label{eq:MAPE}
\end{equation}
evaluated for all response components. The errors are also similarly evaluated for the standard deviations (STD) of the response components. In conclusion, the variability in the observed data is more accurately captured by the output obtained using the transformation-based formulation.
\begin{table}[h!]
\centering
\tabcolsep=3pt
\begin{tabular}{c|rr|rr|}
 & \multicolumn{2}{c|}{MAE [-]} & \multicolumn{2}{c|}{MAPE [\%]} \\
 & MEAN & STD & MEAN & STD \\
 \hline 
 Bayes & $0.0045$ & $0.0039$ & $0.84$ & $15.80$\\
 Transf. & $0.0080$ & $0.0008$ & $1.67$ & $3.44$\\
 \hline
\end{tabular}
\caption{Mean absolute errors of predictive mean values and standard deviations for all response components.}
\label{tab:Ver_comparison_errors}
\end{table}

\subsection{Validation}
\label{validation}

\begin{figure*}
\centering
\includegraphics[width=1.0\linewidth]{ 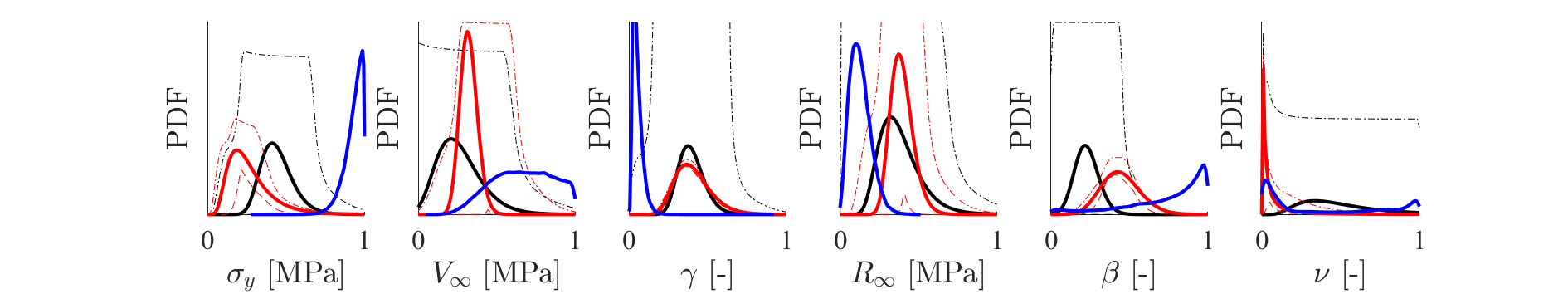} \\
\caption{Hierarchical Bayes: Prior (black) and identified (red) parameters' marginal PDFs corresponding to hyperparameters' mean of prior and mode of posterior (solid lines) accompanied by bounds corresponding to $90$ \% of  hyperparameters' values (dashed lines), transformation: identified (blue) parameters' marginal PDFs.}
\label{fig:kalibrace_parametry}
\end{figure*}

The model calibration based on the real experimental data follows. The results of the model calibration are presented in the same manner as the verification results. The Figure \ref{fig:kalibrace_parametry} shows the identified mean parameters' PDFs with depicted epistemic uncertainties for Bayesian identification and the identified one-dimensional parameters' PDF compared to the prior feasible ranges for transformation-based identification.
Now, it is not possible to compare the results on a~level of model parameters' values. The epistemic uncertainties in Bayesian formulation are reduced the most in the case of the parameter $\gamma$ again due to the experimental setup. Nevertheless, it is important to highlight that the distributions obtained for the two problem formulations differ significantly.

\begin{figure*}
\centering
\includegraphics[width=0.9\linewidth]{ 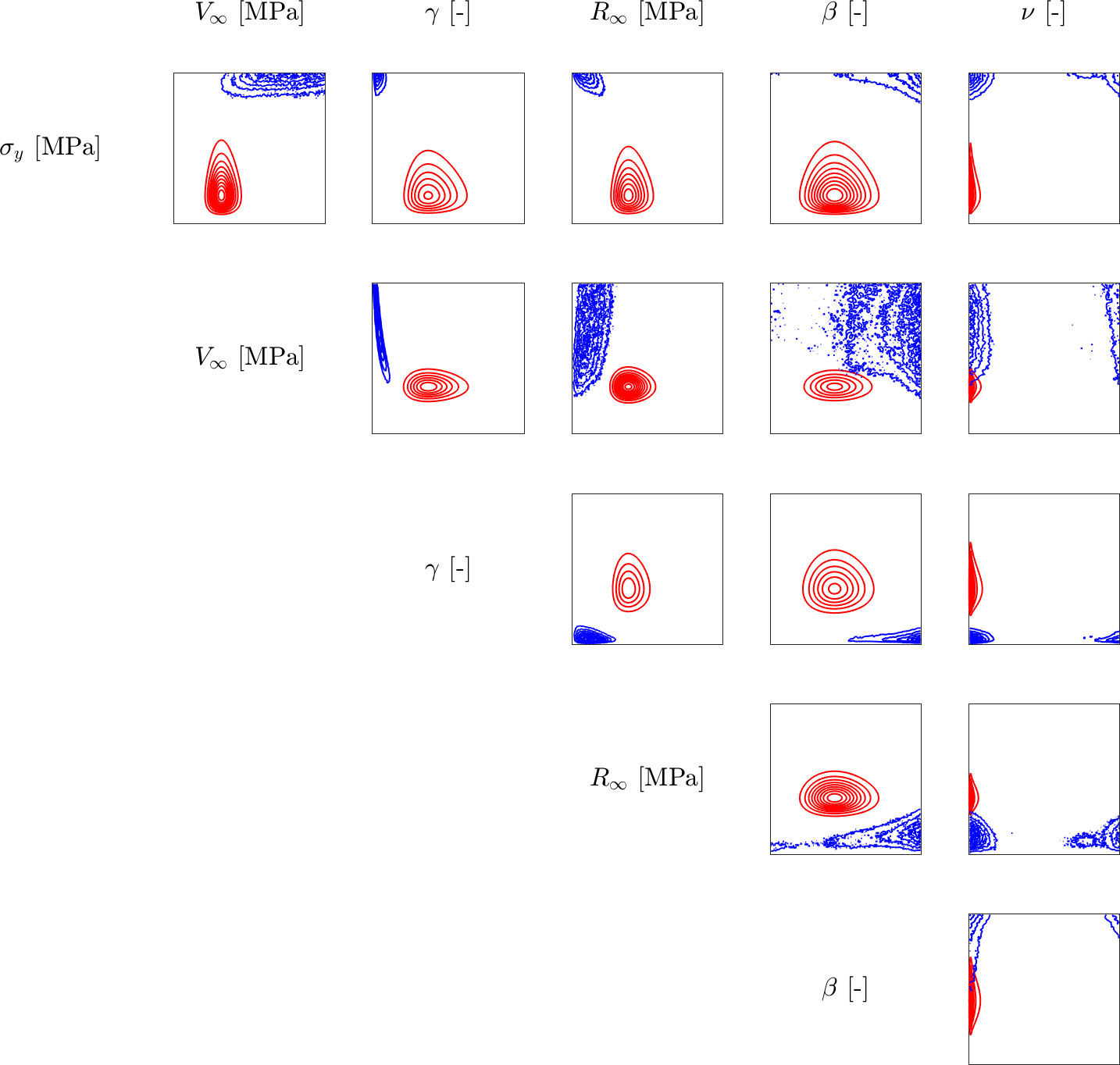}
\caption{Identified parameters' 2D marginal PDFs for Bayesian (red) and transformation-based (blue) formulation.}
\label{fig:kalibrace_parametry_2d}
\end{figure*}

To illustrate the results in more detail, the identified two-dimensional parameters' PDFs obtained from the transformation-based formulation are shown in Figure~\ref{fig:kalibrace_parametry_2d}.  In contrast to the Bayesian formulation, which assumes independent log-normal structural priors, the transformation-based procedure allows to identify an arbitrary joint distribution of parameters. The identified marginal PDFs are evidently more complex and highly dependent. This liberty of transformation-based formulation played here a~crucial role and allowed to provide significantly better results than Bayesian formulation as shown below.

\begin{figure*}
\centering
\begin{tabular}{cc}
\includegraphics[width=0.48\linewidth]{ 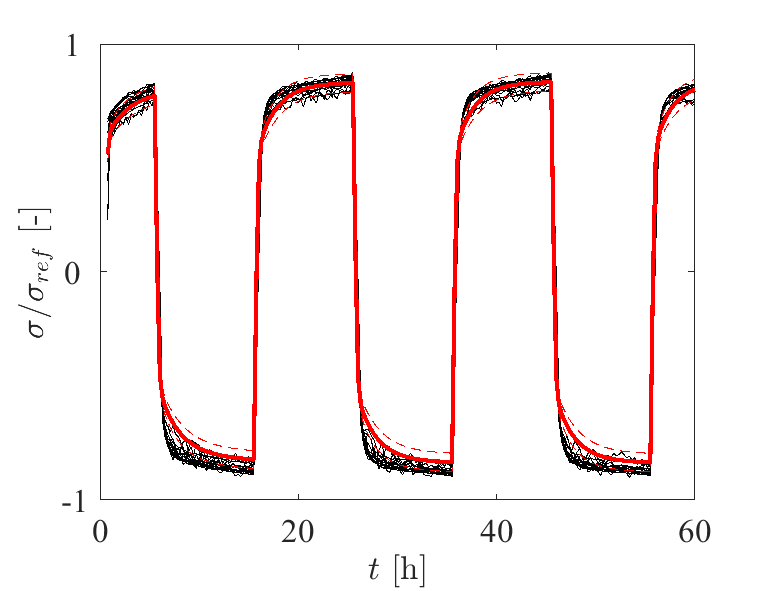} &
\includegraphics[width=0.48\linewidth]{ 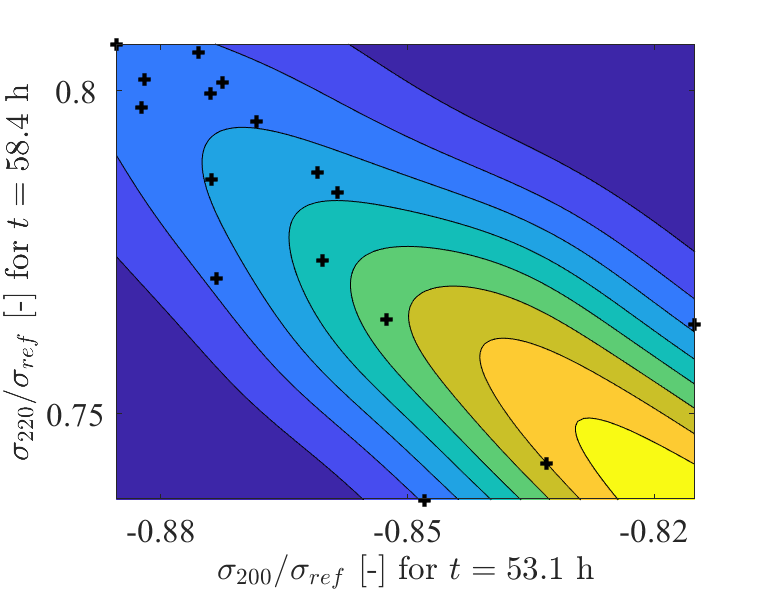} 
\\
\includegraphics[width=0.48\linewidth]{ 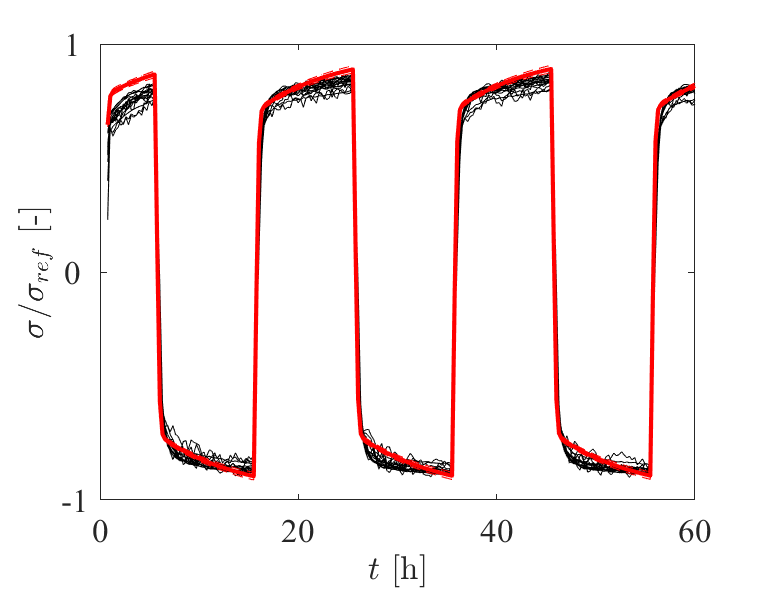} &
\includegraphics[width=0.48\linewidth]{ 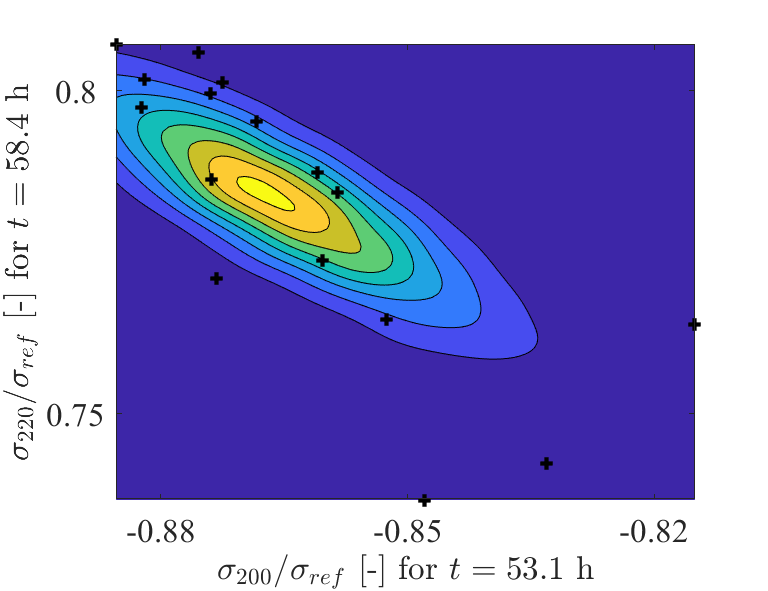} 
\\
(a) & (b)\\
\end{tabular}
\caption{Comparison of experimental observations (black) with model responses (red) corresponding to the identified parameters' distribution in a form of (a) mean value $\pm$ standard deviation, (b) isolines of joint PDF of model responses for $t = \R{53.1}$ h and $t = 58.4$ h.}
\label{fig:validace_napeti}
\end{figure*}

The obtained samples of joint parameters' distribution calibrate the model to predict system behaviour as it is shown in Figure \ref{fig:validace_napeti}. Here, the corresponding model responses are compared to the experimental data as curves Figure \ref{fig:validace_napeti} (a) and the two selected points Figure \ref{fig:validace_napeti} (b).
The first row presents results obtained from Bayesian identification. The standard deviations of particular response components seem to be overestimated. The results demonstrate that the identified mean parameter set is not capable of accurately reproducing the observed mean of selected response components. The results presented in the second row correspond to the transformation-based identification where the predictions seem to achieve a~better agreement.

The numerical results calculated on the basis of $10000$ full numerical model simulations for inputs $\vek{X}$ from identified parameters' distribution for both identification procedures are summarized in Tables \ref{tab:comparison_sigma} and \ref{tab:comparison_errors}. Table~\ref{tab:comparison_sigma} presents the statistical moments of the selected response components $\sigma_{200}$ and $\sigma_{220}$; the results confirm closer agreement between the experimental data and the response of the model calibrated using the transformation-based formulation. 
\begin{table}[h!]
\centering
\tabcolsep=3pt
\begin{tabular}{c|rr|rr|r|}
& \multicolumn{2}{c|}{MEAN} & \multicolumn{2}{c|}{STD} & CORR\\
& $\sigma_{200}$ & $\sigma_{220}$ & $\sigma_{200}$ & $\sigma_{220}$ & $\sigma_{200}, \sigma_{220}$\\
 \hline
 Data & $-0.864$ & $0.782$ & $0.019$ & $0.022$& $-0.775$\\
 Bayes & $-0.820$ & $0.739$ & $0.049$ & $0.049$ & $-0.957$\\
 Transf. & $-0.865$& $0.783$ & $0.015$ & $0.012$ & $-0.872$ \\
 \hline
\end{tabular}
\caption{Comparison of statistical moments of the observed data and the predictive model responses for both identification formulation.}
\label{tab:comparison_sigma}
\end{table}

Table \ref{tab:comparison_errors} presents MAE and MAPE evaluated for all response components. The results obtained for the experimental data are generally worse due to experimental errors, model inadequacies and the lack of knowledge about the true system setting. The validation results indicate that the formulation of the stochastic inversion problem based on transformation of variables provides a more accurate fit.
\begin{table}[h!]
\centering
\tabcolsep=3pt
\begin{tabular}{c|rr|rr|}
 & \multicolumn{2}{c|}{MAE [-]} & \multicolumn{2}{c|}{MAPE [\%]} \\
 & MEAN & STD & MEAN & STD \\
 \hline
Bayes & $0.053$ & $0.027$ & $10.27$ & $120.97$\\
Transf. & $0.045$ & $0.012$ & $10.01$ & $37.85$\\
 \hline
\end{tabular}
\caption{Mean absolute errors of predictive mean values and variances for all response components.}
\label{tab:comparison_errors}
\end{table}

\section{Estimation of parameter distributions from independent tests}
\label{sec:model2}

The second investigated case concentrates on identification of distribution of two damage parameters from two distinct types of destructive experiments. The damage model is governed by the Landgraf - Morrow equation, see~~\cite{Landgraf:1970}, which expresses the relationship between the strain range $\Delta \varepsilon$ and the number of cycles to failure $N_f$ of a~specimen as
\begin{equation}
    N_f = \left( \frac{\Delta\,\varepsilon }{S} \right)^{-s} \, .
\end{equation}
The material parameters to be identified are the fatigue ductility coefficient $S$ [-] and the fatigue ductility exponent $s$ [-]. Based on expert knowledge, the parameters are limited to feasible intervals $S \in [0.1; 1.2]$ and $s \in [1.0 ; 2.8]$.

\begin{figure}[h!]
\begin{center}
\includegraphics*[width=1\linewidth,keepaspectratio]{ 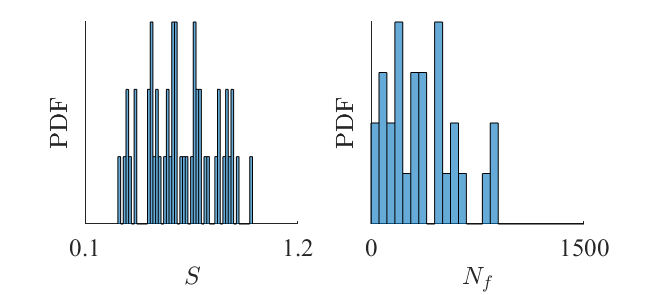}
\caption{Histograms of synthetic observations.}
\label{fig:Damage_data}
\end{center}
\end{figure}

To identify the damage parameters, two types of destructive experiments are considered, and only synthetic data are used in this study. The first experiment comprises $50$ repetitions of a tensile test, where the measured quantity is the strain at rupture, which directly corresponds to the parameter~$S$
The second experiment consists of 30 repetitions of a cyclic loading test, in which the number of cycles to failure $N_f$ is measured under a fixed strain range $\Delta \varepsilon = 0.03$. The synthetic data are generated for model inputs sampled from independent marginal Beta distributions with identical hyperparameters $\alpha = 2$ and $\beta = 2$, scaled to the respective feasible intervals of the damage para\-meters. The histograms of the resulting synthetic data are shown in Figure~\ref{fig:Damage_data}. 

\subsection{Bayesian hierarchical model}
\label{sec:Damage-Bayes}

In Bayesian formulation, the Beta distribution scaled to feasible intervals is considered also as a prior distribution for the damage parameters. Here its hyperparameters $\alpha$ and $\beta$ are unknown with uniform prior
\begin{equation}
    \alpha, \beta \sim U(0.1,15) .
\end{equation}
To capture an incidental dependency between the parameters, the correlation coefficient $\rho$ is introduced with a uniform hyperprior
\begin{equation}
    \rho \sim U(-1,1) .
\end{equation}
The joint bivariate probability distribution of para\-meters is described with the help of a Gaussian copula \cite{Nelsen:2006} according to Sklar's theorem \cite{Sklar:1959} as
\begin{equation}
    f_{S,s}(S, s) = c_\rho \cdot f_S(S) \cdot f_s(s) \, , \\
 \end{equation}   
where \R{$f_S(S)$ and $f_s(s)$ are the PDFs of the Beta distributions with hyperparameters $(\alpha_S, \beta_S)$ and $(\alpha_s, \beta_s)$, respectively.} The density of the Gaussian copula
\begin{equation}
    c_\rho = \frac{1}{\sqrt{1 - \rho^2}} \exp\left( 
        -\frac{1}{2(1 - \rho^2)} 
        \left[ z_1^2 + z_2^2 - 2\rho z_1 z_2 \right] 
    \right) \, 
\end{equation} 
introduces the dependence via the standard normal quantile transformation
\begin{equation}
\begin{tabular}{c}
    $z_1 = \Phi^{-1}(F_S(S))\,$, \\
    $z_2 = \Phi^{-1}(F_s(s))\,$, \\
\end{tabular}
\end{equation} 
where \(\Phi^{-1}\) denotes the inverse cumulative distribution function of the standard normal distribution, $F_S(S)$ and $F_s(s)$ represent cumulative distribution functions of the marginal Beta distributions.

The likelihood function (\ref{eq:likelihood}) involves the data from the tensile tests ($n_1=50$) with the corresponding i.i.d. experimental errors $E_1 \sim N(0, 0.1^2)$ for each specimen and from the cyclic tests ($n_2=30$) with the i.i.d. experimental errors $E_2 \sim N(0, 0.8^2)$ for each specimen.

The posterior distribution comprises $160$ material parameters (one pair corresponding to each of $80$ observations) and $5$ hyperparameters defining the aleatory uncertainty in the examined material. The posterior MCMC samples are obtained by Metropolis algorithm with a Gaussian random step with standard deviation set up according to optimal acceptance rate ($20-50$~\%). 

\begin{figure}[h!]
\begin{center}
\begin{tabular}{c}
\includegraphics*[width=1\linewidth]{ 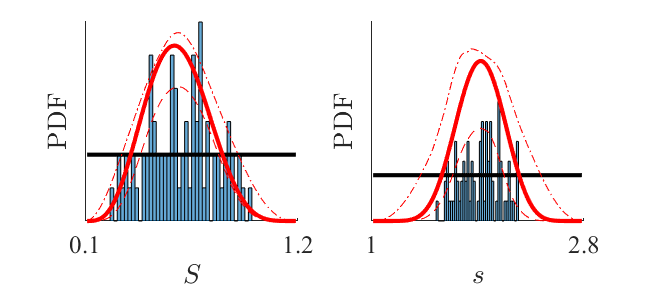}\\
\includegraphics*[width=0.48\linewidth]{ 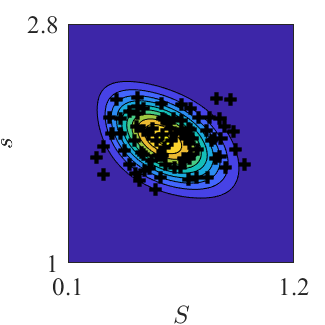}\\
\end{tabular}
\end{center}
\caption{Bayes: Histograms of synthetic experimental inputs, prior (black) and posterior (red) parameters' PDFs corresponding to MAP hyperparameters accompanied by bounds corresponding to $90$ \% of  predicted PDF values. Isolines of identified joint parameters' PDF in comparison with all $80$ synthetic experimental inputs.} 
\label{fig:Damage_pdfs_Bayes}
\end{figure}

Figure \ref{fig:Damage_pdfs_Bayes} compares the identified
\begin{figure*}[b]
\begin{center}
\includegraphics*[width=1\linewidth,keepaspectratio]{ 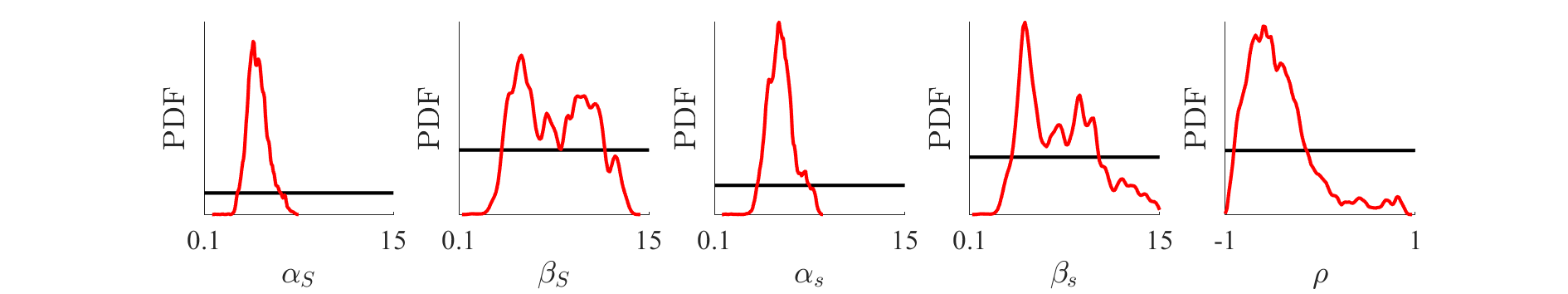}
\end{center}
\caption{Bayes: Identified (red) marginal distributions of hyperparameters in comparison to their hyperprior (black).} 
\label{fig:Damage_pdfs_Bayes_hyperpar}
\end{figure*}
marginal distributions of the material parameters to the histograms of synthetic data inputs.
The red solid lines correspond to the marginal beta PDFs of the damage parameters for the identified MAP estimates of hyperparameters $\alpha$ and $\beta$, i.e. the best estimate of the damage parameters' distribution (representing the aleatory uncertainties). The dashed red lines form an envelope of the damage parameters' PDFs corresponding to the $5$\% and $95$\% quantiles of their values (estimated from $1.000$ samples for each MCMC sample of the hyperparameters) representing the epistemic uncertainties in the shape of the estimated PDF of the damage parameters. Both PDFs interleave the data well, the fatigue ductility exponent $s$ is identified with a higher level of epistemic uncertainty. The joint distribution is shown as isolines of the Gaussian copula with MAP Beta marginals and the identified mean value of the correlation coefficient, which points out a negative dependence between the parameters. The identified marginal distributions of the hyperparameters are depicted in Figure \ref{fig:Damage_pdfs_Bayes_hyperpar}. The lowest level of epistemic uncertainty corresponds to the hyperparameters $\alpha_S$ and $\alpha_s$. \R{Some posterior distributions of the hyperparameters exhibit multimodality, reflecting multiple hyperparameter combinations yielding similar likelihood values. After discarding burn-in, convergence was assessed using running means and effective sample size estimates, indicating stable posterior summaries and several hundred effectively independent samples per hyperparameter.}

\begin{figure}[h!]
\begin{center}
\begin{tabular}{c}
\includegraphics*[width=1\linewidth,keepaspectratio]{ 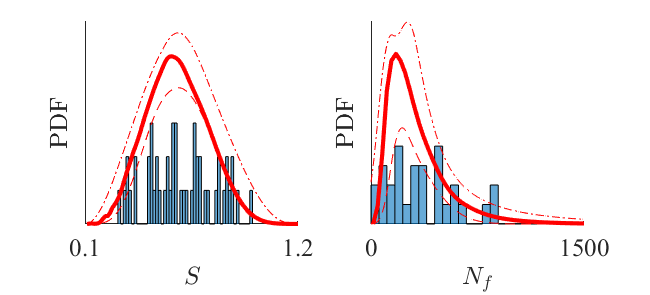}\\
\includegraphics*[width=0.5\linewidth,keepaspectratio]{ 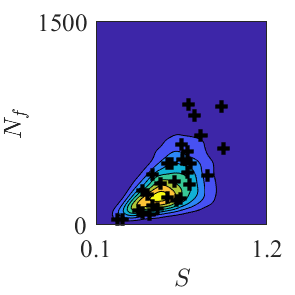}\\
\end{tabular}
\end{center}
\caption{Bayes: Histograms of synthetic data and model predictions for identified MAP damage parameters' PDF accompanied by bounds corresponding to $90$ \% of predicted PDF values. Isolines of predictions' PDF in comparison with $30$ synthetic experimental outputs from cycling loading test.} 
\label{fig:Damage_pred_Bayes}
\end{figure}

The predictive ability of the posterior PDF appears to be satisfactory; see Figure \ref{fig:Damage_pred_Bayes}
comparing the histograms of synthetic data and the distributions of model responses corresponding to the identified MAP parameters' PDF.

\subsection{Transformation of variables}
\label{sec:Damage-transform}

\begin{figure}[h!]
\begin{center}
\begin{tabular}{c}
\includegraphics*[width=1\linewidth,keepaspectratio]{ 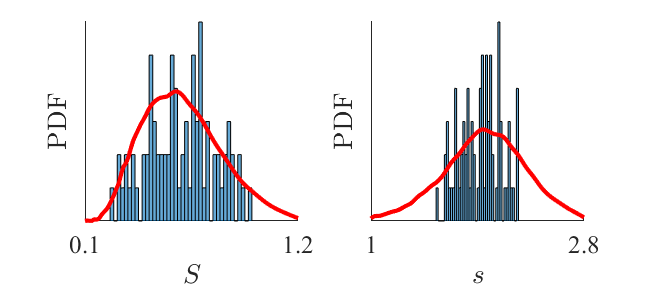}\\
\includegraphics*[width=0.5\linewidth,keepaspectratio]{ 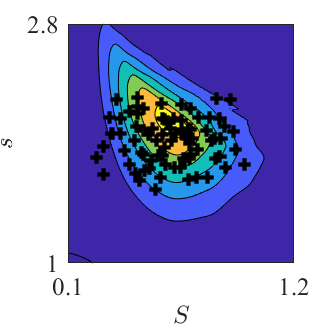}\\
\end{tabular}
\end{center}
\caption{Transformation: Histograms of synthetic experimental inputs and identified marginal PDFs of damage parameters. Isolines of identified joint parameters' PDF in comparison with all $80$ synthetic experimental inputs.} 
\label{fig:Damage_pdfs_PCA}
\end{figure}

Data observation from two distinct types of experiments on different specimens leads to missing information about the inherent correlation in the data. Therefore, this correlation is estimated approximately from model simulations performed for samples of material parameters generated from prescribed feasible intervals. The estimated correlation is combined with the estimated marginal distributions of the data based on a normal kernel function to generate new synthetic data with desired correlation. These are used to build the approximation of their joint probability density function, which is applied for the transformation of variables according to (\ref{eq:transform_PCA}) to obtain the distribution of material parameters depicted in Figu\-re~\ref{fig:Damage_pdfs_PCA}.

\begin{table*}[b]
\centering
\tabcolsep=2pt
\begin{tabular}{c|c|c|c|c|c|c|c|c|c|c|c|c|c|c|c|c|}
& \multicolumn{6}{c|}{MEAN} & \multicolumn{6}{c|}{STD} & \multicolumn{4}{c|}{CORR}\\
\hline
 Parameter & \multicolumn{2}{c|}{$S$} & \multicolumn{2}{c|}{$s$} & \multicolumn{2}{c|}{$N_f$} & \multicolumn{2}{c|}{$S$} & \multicolumn{2}{c|}{$s$} & \multicolumn{2}{c|}{$N_f$} & \multicolumn{2}{c|}{$S,s$} & \multicolumn{2}{c|}{$S,N_f$}\\
 & $\mu$ & $\sigma$ & $\mu$ & $\sigma$ & $\mu$ & $\sigma$ & $\mu$ & $\sigma$ & $\mu$ & $\sigma$ & $\mu$ & $\sigma$ & $\mu$ & $\sigma$ & $\mu$ & $\sigma$ \\
 \hline
 Data & $0.595$& & $1.924$& & $406$& & $0.175$& & $0.174$& & $337$& & $-0.011$& & $0.836$ &\\
 Bayes & $0.595$ & $0.03$ & $1.927$ & $0.07$ & $366$ & $78$ & $0.170$ & $0.02$ & $0.240$ & $0.04$ & $289$ & $142$ & $-0.431$ & $0.36$ & $0.315$ & $0.20$\\
 Transf. & $0.602$& & $1.962$& & $451$& & $0.199$& & $0.340$& & $432$& & $-0.400$& & $0.334$& \\
 \hline
\end{tabular}
\caption{Comparison of statistical moments of the synthetic inputs and observations with the identified parameters' PDFs and the predictive model responses for both identification formulations.}
\label{tab:comparison_damage}
\end{table*}

\begin{figure}[h!]
\begin{center}
\begin{tabular}{cc}
\includegraphics*[width=1\linewidth,keepaspectratio]{ 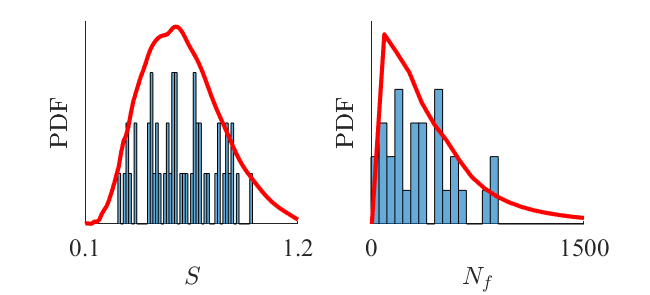}\\
\includegraphics*[width=0.5\linewidth,keepaspectratio]{ 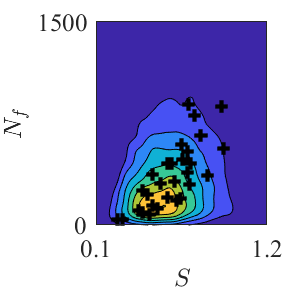}\\
\end{tabular}
\end{center}
\caption{Transformation: Histograms of synthetic data and model predictions for identified PDF of damage parameters. Isolines of predictions' PDF in comparison with $30$ synthetic experimental outputs from cycling loading test.} 
\label{fig:Damage_pred_PCA}
\end{figure}

The identification of the fatigue ductility coefficient $S$ is essentially perfect, while the variance of the fatigue ductility exponent $s$ is overestimated. The differen\-ce obtained in the variance of $s$ can be caused by an inaccurate estimation of the joint distribution of both data components $S$ and $N_f$, which here is principally unknown. Therefore, this joint distribution is approximated at two levels. First, both components are modelled as simply linearly dependent with estimated correlation. Second, their correlation is only roughly estimated on the basis of other simulated data sets obtained for parameters distributed in feasible intervals. However, according to Figure~\ref{fig:Damage_pred_PCA} the known marginal distributions of both data components are fitted with high accuracy. As a possible simple extension of this work, the applied identification steps can be repeated sequentially. A new estimate of data correlation can be obtained based on new simulations from the identified distribution of material parameters instead of the original feasible intervals and further app\-lied to gain a new updated estimation of parameter distributions. 

\subsection{Comparison}
The results of the damage model calibration obtained for both identification formulations are summarized in Table \ref{tab:comparison_damage}. Specifically, the statistical moments (mean values (MEAN), standard deviations (STD) and corre\-lations (CORR)) of the identified aleatory uncertainties in the parameters as well as in the calibrated model predictions are compared to the synthetic data (values corresponding to the damage parameters $S$ and $s$ are calculated from $80$ samples, values corresponding to the observable $N_f$ from $30$ samples). 
 
Both identification formulations achieve good agreement between the identified and synthetic inputs in terms of the parameters' mean values (maximum error: \(2\%\) of the feasible interval). However, the Bayesian formulation estimates the parameters' variability more accurately than the transformation-based approach, which tends to overestimate their standard deviations. The mean and standard deviation of the model res\-ponse \(N_f\) are predicted with lower accuracy, as evidenced by the relatively large epistemic uncertainty \(\sigma\) estimated in Bayesian identification. In general, when information on data correlation is lacking due to the experimental setup, the transformation-based formulation is not preferred, even if additional information from model simulations is available. In cases with limited data, the Bayesian formulation—capable of quantifying epistemic uncertainties—is recommended.

\section{Conclusions}
\label{sec:conclusions}

The inherent variability in the properties of heterogeneous materials can be formulated as a stochastic inverse problem, where the solution is a probabi\-lity distri\-bution of material parameters. These para\-meters, when used in numerical model simulations, should produce output distributions consistent with those of the observed experimental data.

This paper addresses the identification of aleatory uncertainties associated with randomicity of material parameters due to material heterogeneity. In parti\-cular, the kinematic and isotropic hardening material model for a copper alloy is successfully calibrated from the experimental data set comprising $16$ repetitions of a~strain-controlled cyclic loading test at a high temperature. Two fundamentally different formulations of the identification problem are proposed and compared, including a case where data originate from two distinct sample groups, each subjected to a different destructive experiment, where no single experiment provides sufficient information to fully identify all model parameters.

The first formulation of the stochastic inversion problem is based on the multilevel setting of Bayesian inference, which provides an elegant solution for stochastic inversion problems. This efficient tool enab\-les one to separate epistemic uncertainties and inhe\-rent variability, which is fundamental for successful parameter identification of heterogeneous material model from indirect noisy measurements. The formulation is particularly beneficial in cases with limited information, such as data obtained from different types of destructive experiments. The major disadvantage is the requirement of assuming some specific type of parameters' PDF which affects the result essentially, and it can be problematic, for example, in the case of phenomenological parameters. The formulated posterior probability distribution can be gained in the form of independent samples by the Markov chain Monte Carlo method; however, the computational demands increase significantly with a number of measurements, so this approach is suitable especially for small experimental data sets.

The underlying benefits of the transformation-based formulation are reducing assumptions about the unknown parameter PDF structure and efficient experi\-mental data processing. The parameters are const\-rained solely by their feasible domain, within which the numerical model is defined. Simplifying assumptions are introduced only for the probability density function of the data, which is arguably less misleading, as samples from this distribution are actually observed. The formulated joint distribution of the parameters can be gained in the form of independent samples by MCMC, too, but the dimension of the searched space does not change with additional observation as it is in the first formulation. Polynomial approximation of the calibrated numerical model is applied to dimi\-nish computational demands, and moreover it enables to compute the transformation constant analytically. This approach suffers from the disadvantage of quantifying all present uncertainties without distinguishing between aleatory and epistemic ones. \R{An extension of the method that would explicitly incorporate a known measurement error model, or estimate it jointly with the material parameters, represents a natural direction for future work.}

\section*{Acknowledgements}
This work was supported by the European Union through the Operational Programme Jan Amos Komensk\'{y} under project INODIN (CZ.02.01.01/00/23\_020/0008487).

\end{document}